\documentclass[aps,prx,noshowpacs,twocolumn,nofootinbib]{revtex4-2}
\usepackage[utf8]{inputenc}
\usepackage{xcolor}
\usepackage{graphics,graphicx,epsfig,multirow}
\usepackage{amssymb,amsfonts,amsmath,bm}
\usepackage{algorithm,algpseudocode}
\usepackage{ifthen}

\usepackage{hyperref}

\usepackage{dsfont}
\renewcommand{\paragraph}[2][.]{\noindent {\bf #2#1}}		

\newcommand{\EQ}{\begin{equation}}
\newcommand{\EE}{\end{equation}}
\newcommand{\EQA}{\begin{eqnarray}}
\newcommand{\EEA}{\end{eqnarray}}

\newcommand{\ts}{{\text{ts}}}
\newcommand{\tv}{{\text{tv}}}

\begin{document}

\title{Design of an optimal combination therapy with broadly neutralizing antibodies to suppress HIV-1}
\author{Colin LaMont}
\affiliation{Max Planck Institute for Dynamics and Self-organization, Am Fa\ss berg 17, 37077 G\"ottingen, Germany}
\author{Jakub Otwinowski}
\altaffiliation{Current address: Dyno Therapeutics, 1 Kendall Square, Cambridge, MA 02139}
\affiliation{Max Planck Institute for Dynamics and Self-organization, Am Fa\ss berg 17, 37077 G\"ottingen, Germany}
\author{Kanika Vanshylla}
\affiliation{Laboratory of Experimental Immunology, Institute of Virology, Faculty of Medicine and University Hospital Cologne, University of Cologne, 50931 Cologne, Germany}
\author{Henning Gruell}
\affiliation{Laboratory of Experimental Immunology, Institute of Virology, Faculty of Medicine and University Hospital Cologne, University of Cologne, 50931 Cologne, Germany}
\author{Florian Klein}
\affiliation{Laboratory of Experimental Immunology, Institute of Virology, Faculty of Medicine and University Hospital Cologne, University of Cologne, 50931 Cologne, Germany}
\affiliation{Partner Site Bonn-Cologne, German Center for Infection Research, 50931 Cologne, Germany}
\author{Armita Nourmohammad}
\email{Correspondence should be addressed to Armita Nourmohammad: armita@uw.edu}
\affiliation{Department of Physics, University of Washington, 3910 15th Ave Northeast, Seattle, WA 98195, USA}
\affiliation{Max Planck Institute for Dynamics and Self-organization, Am Fa\ss berg 17, 37077 G\"ottingen, Germany}
\affiliation{Fred Hutchinson Cancer Research Center, 1100 Fairview ave N, Seattle, WA 98109, USA}

\date{\today} 
\begin{abstract}
\noindent 
Broadly neutralizing antibodies (bNAbs) are promising targets for vaccination and therapy against HIV. Passive infusions of bNAbs have shown promise in clinical trials as a potential alternative for anti-retroviral therapy. A key challenge for the potential clinical application of bnAbs is the suppression of viral escape, which is more effectively achieved with a combination of bNAbs. However, identifying an optimal bNAb cocktail is combinatorially complex. Here, we propose a computational approach to predict the efficacy of a bNAb therapy trial based on the population genetics of HIV escape, which we parametrize using high-throughput HIV sequence data from a cohort of untreated bNAb-naive patients. By quantifying the mutational target size and the fitness cost of HIV-1 escape from bNAbs, we reliably predict the distribution of rebound times in three clinical trials. Importantly, we show that early rebounds are dominated by the pre-treatment standing variation of HIV-1 populations, rather than spontaneous mutations during treatment. Lastly, we show that a cocktail of three bNAbs is necessary to suppress the chances of viral escape below 1\%, and we predict the optimal composition of such a bNAb cocktail. Our results offer a rational design for bNAb therapy against HIV-1, and more generally show how genetic data could be used to predict treatment outcomes and design new approaches to pathogenic control.
\end{abstract}
\maketitle

\section{Introduction}
Recent discoveries of highly potent broadly neutralizing antibodies (bNAbs) provide new opportunities to successfully prevent, treat, and potentially cure infections from evolving viruses such as HIV-1~\cite{Walker:2009cd,Walker:2011ew,Liao:2013gs,Mouquet:2013he,Klein:2013eb,Kwong:2013ia,Caskey:2015hm,Caskey:2017el,bar-onSafetyAntiviralActivity2018,sokRecentProgressBroadly2018,zwickBroadlyNeutralizingAntibodies2001,burtonBroadlyNeutralizingAntibodies2012}, influenza~\cite{Sparrow:2016ew}, and the Dengue virus~\cite{Ekiert:2012gr,Durham:2019fq}. 
bNAbs target vulnerable regions of a virus, such as the CD4 binding site of HIV {\em env} protein,  where escape mutations {can be} costly for the virus~\cite{Walker:2009cd,Chen:2009ig,Zhou:2010gx,Walker:2011ew,Liao:2013gs,West:2014jq,Burton:2016ek}. As a result, eliciting bNAbs is  the goal of a universal vaccine design against the otherwise rapidly evolving HIV-1. Apart from vaccination, bNAbs  can also offer significant advances in  therapy against both HIV and influenza~\cite{West:2014jq,Caskey:2016kk,Gruell:2018ixa,Durham:2019fq}. Specifically, augmenting current anti-retroviral therapy (ART) drugs with bNAbs may provide the next generation of HIV therapies~\cite{Horwitz:2013gb,Gruell:2018ixa}. 

Recent studies have used bNAb therapies to curb infections by the Simian immunodeficiency virus (SHIV)  in non-human primates~\cite{Shingai:2013kn,Barouch:2013gn,Julg:2017kg}, and  HIV-1 infections in human clinical trials~\cite{Caskey:2015hm,Bar:2016hg,Caskey:2017el,bar-onSafetyAntiviralActivity2018}. Monotherapy  trials with  potent bNAbs, including  3BNC117~\cite{Caskey:2015hm}, VRC01~\cite{Bar:2016hg}, and 10-1074~\cite{Caskey:2017el}  indicate that administering bNAbs is safe and can suppress viral load in patients.  Nonetheless, in each trial, escape mutants emerge resulting in a viral rebound after about 20 days past infusion of  the bNAb. However, in trials that administered a combination of 10-1074 and 3BNC117, viral rebound was substantially suppressed~\cite{Shingai:2013kn,bar-onSafetyAntiviralActivity2018}. Combination therapy has been repeatedly used against many infectious agents,  including current HIV ART cocktails and combination antibiotic treatments against Tuberculosis~\cite{Lienhardt:2012fl}. The principle behind combination therapy is clear: It is harder for a pathogen population to acquire resistance  against multiple treatment targets simultaneously than to acquiring resistance against each target separately.

{Prior work has focused on optimizing combination therapy of bNAbs for breadth and potency \cite{yuPredictingBroadlyNeutralizing,waghOptimalCombinationsBroadly2016a} without considering viral dynamics.}   Theoretical approaches have been used to model the dynamics of viremia in patients following passive infusion of bNAbs~\cite{Lu:2016id,Reeves:2020ca,Saha:2020fd,meijersPredictingVivoEscape2021}. However the predictive power of models relying on trial data is limited by the small number of individuals enrolled in these trials, and increasing the size of a trial may be impractical. One can view the outcome of these trials in the context of an evolutionary competition among susceptible and resistant strains with different probabilities to emerge and establish in an HIV population within a patient, due to their different differential fitness effects  in the presence or absence of a bNAb. Studies of population genetics of HIV have found rapid intra-patient evolution and turnover of the virus~\cite{Lemey:2006wb,Zanini:2015gg} and have indicated that the efficacy of drugs in anti-retroviral therapy can severely impact the mode of viral evolution and escape~\cite{Feder:2016bc}. 
Despite the  complex evolutionary dynamics of HIV-1 within patients due to individualized immune pressure~\cite{Nourmohammad:2019ij}, genetic linkage~\cite{Zanini:2015gg}, recombination~\cite{Neher:2010dw,Zanini:2015gg}, and epistasis between loci~\cite{Bonhoeffer:2004cf,Zhang:2020bs}, the genetic composition of a population can still provide valuable information about the evolutionary significance of specific mutations, especially  in highly vulnerable regions of the virus. For example, analysis of genomic covariation in the Gag protein  of HIV-1 has been successful in predicting fitness effect of mutations in relatively conserved regions of the virus, which could inform the design of rational T-cell therapies that  target these vulnerable regions~\cite{Ferguson:2013kb}.   

Here, we present a statistical inference framework that uses the high throughput longitudinal survey of genetic data collected from 11 ART-naive patients over about 10 years of infection~\cite{Zanini:2015gg} to characterize the evolutionary fate of escape mutations and to predict patient outcomes in recent mono- and combination therapy  trials with 10-1074 and 3BNC117 bNAbs~\cite{Caskey:2015hm,Caskey:2017el,bar-onSafetyAntiviralActivity2018}. Using the accumulated intra-patient genetic variation from deep sequencing of HIV-1 populations in ART--naive patients~\cite{Zanini:2015gg}, we can estimate the diversity and the fitness effects of mutations at sites mediating escape. These variables parametrize our  individual-based model  for viral dynamics to characterize the expected path for a potential escape of HIV-1 populations in response to bNAb therapies in patients enrolled in the clinical trials. Our analysis accurately predicts the distribution of viral rebound times in response to passive  bNAb  infusions~\cite{Caskey:2015hm,Caskey:2017el,bar-onSafetyAntiviralActivity2018}, measuring the efficacy of these clinical trials. 

Our prediction for the viral rebound time in response to a bNAb is done based on the inferred genetics parameters from the deep sequencing of HIV-1 populations in a separate cohort of ART-naive patients. Therefore, we use our approach to  assess a broader panel of nine bNAbs, for which escape sites can be identified from prior deep mutational scanning experiments~\cite{Dingens:2019fd}, to characterize the therapeutic efficacy of each of these bNAbs and to propose optimal combination therapies that can efficiently curb an HIV infection. Our results showcase how the wealth of genetic data can be leveraged to guide rational therapy approaches against HIV. Importantly, this approach is potentially applicable to therapy designs against other evolving pathogens, such as resistant bacteria or cancer. 

\section{Model}
\begin{figure*} [t!]
    \centering
	\includegraphics[width=0.8\textwidth]{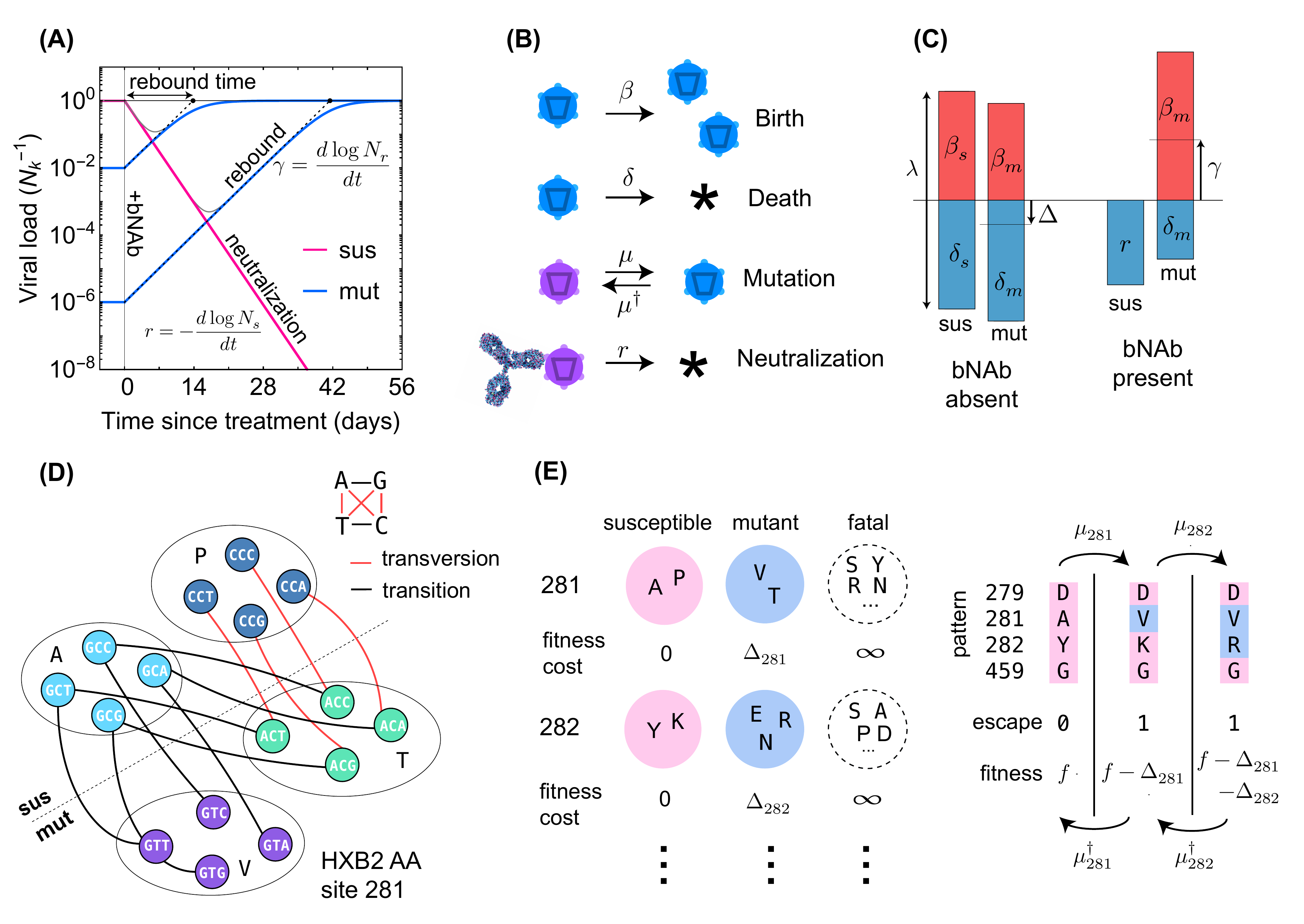}
    \caption{
    {\bf Schematics for the evolutionary dynamics of viral rebound.}
{\bf (A) }
	The viral dynamics after the initiation of a treatment with bNAb infusion ($t = 0$) is determined by two competing processes. 
	Susceptible strains (sus) undergo exponential decay (red line) with decay rate given by $r$, 
		while the resistant mutants (mut) undergo logistic growth back up to the carrying capacity ($N_k$) of the patient.
	In the deterministic limit (eq.~\ref{eq:logisticpiecewise}), the rebound time is linearly related to the log-frequency of the mutant fraction.
{\bf (B) }
	The schematic shows the four stochastic processes of birth, death, mutation, and neutralization with their respective rates for susceptible (purple) and resistant (blue) variants. These processes define the evolution of a  viral population. Note that both the susceptible and the resistant variants are subject to birth and death with their respective rates.
{\bf (C) }
	The birth and death rates can be visualized as a region of size $\lambda = \beta+\delta$ which is partitioned into birth and death events.
	In the absence of antibodies, the susceptible population has balanced birth and death rates, $\beta_s=\delta_s$, while
		the resistant population has a negative net birth rate equal to the fitness difference $\Delta =\delta_m-\beta_m$.
	After introduction of the antibody, the susceptible population decays at rate $r$, and without competition from the susceptible population, the resistant population grows at the free growth-rate $\gamma$.
{\bf (D) }
	Mutational target size is inferred {\it a priori} from the genotype-phenotype mapping, 
		which can be visualized as a bipartite graph.
	The nodes correspond to codons, while the
		edges are the mutations which link one codon to another, weighted according to the respective mutation rates.
	The average edge weight from codons of susceptible variants to the escape  mutants determines the rate of escape mutations $\mu$. 
	Mutations can be divided into two types: transitions (black) are within-class, and transversions (red) are out of class nucleotide changes.
	Transitions occur at about  $8$ times higher rate than transversions (Fig.~\ref{Fig2}).
	 {\bf (E) }
	A coarse grained fitness and mutation model for two of the escape sites (281 and 282) against antibody 3BNC117 are shown.
	Left: At each escape mediating site, amino acids fall into one of three groups: (i) susceptible (wild-type), (ii) escape mutant, and (iii) fatal.
	For an escape-class amino acid at site $i$ the virus incurs a fitness cost $\Delta_i$, and these costs are additive across sites.
	Right: Mutations at a given site $i$ occur with (independent) forward $\mu_i$ and backward $\mu^\dagger_i$  rates which govern the substitution events between amino acid classes.}
   \label{Fig1}
\end{figure*}
\subsection*{HIV response to therapy}
After infusion of bNAbs in a patient, the antibodies bind and neutralize the susceptible strains of HIV. The neutralized subpopulation of HIV no longer infects T-cells, 
	and the plasma RNA copy-number associated with this neutralized population decays. The dynamics of viremia in HIV patients off ART following a bNAb therapy with 3BNC117~\cite{Caskey:2015hm}, 10-1074~\cite{Caskey:2017el}, and their combination~\cite{bar-onSafetyAntiviralActivity2018} are shown in Figs.~S1-S3.
With competition of the neutralized strains removed,	the resistant subpopulation grows until the viral load typically recovers to a level close to the pretreatment state (i.e., the carrying capacity); see Fig.~\ref{Fig1}A. The time it takes for the viral load to recover is the {\em rebound time}-- a key quantity that characterizes treatment efficacy within a patient. Although the details of the viremia dynamics, especially at beginning and at the end of the therapy,  may be complex~\cite{Lu:2016id,Reeves:2020ca,Saha:2020fd,meijersPredictingVivoEscape2021}, the rebound time  can be approximately modeled using a logistic growth after bNAb infusion ($t>0$),
\begin{align}
\label{eq:logisticpiecewise}
N(t) = 
\begin{cases}
N_k 
	& t \leq 0 \\
(1-x) N_k e^{-r t} + \frac{N_k}{1+ \frac{1-x}{x}e^{- \gamma t}}
	& t>0
\end{cases}
\end{align}
with the initial condition set for pre-treatment fraction of resistant subpopulation $x=  N_r(0)/ (N_r(0) + N_s (0))$, where $N_r(0)$ and $N_s(0)$ denote the size of resistance and susceptible subpopulations at time $t=0$, respectively. Here, $\gamma$ is the growth rate of the resistant population, $r$ is the neutralization rate impacting the susceptible subpopulation, and $N_k$ is the carrying capacity (Fig.~\ref{Fig1}A, Methods).  In our analysis, we set $\gamma = 1/3 \text{ days}^{-1}$ or a doubling time of $\sim 2$ days, which is the characteristic of HIV growth in patients~\cite{Perelson:1996hv}. We infer the neutralization rate $r$ as a global parameter for each trial, since it depends on the neutralization efficacy of a bNAb at the concentration used in the trial. We will infer the patient-specific pre-treatment fraction  of resistant subpopulation $x$, using a population genetics based approach based on which we characterize the mutational target size and selection cost of escape in the absence of a bNAb (see below).  The resulting viremia fits in Figs.~S1-S3 specify the rebound time $T$ in each patient, which in this simple model, is given by $T =-\gamma^{-1} \log x$ (Methods).

The rebound time following passive infusion of  3BNC117~\cite{Caskey:2015hm} and 10-1077~\cite{Caskey:2017el} bNAbs range from 1 to 4 weeks, with a small fraction of patients exceeding the monitoring time window in the studies (late rebounds past 56 days);  see Figs.~S1-S3. The distribution of rebound times summarizes the escape response of the virus to a therapy and directly relates to the distribution for the pre-treatment fraction of resistant variants $P(x)$ across patients $P(T) \sim x^{-1} P(x)$.

\subsection*{Stochastic evolutionary dynamics of HIV subject to bNAb therapy}
The fate of an HIV population subject to bNAb therapy depends on the composition of  the pre-treatment population with resistant and susceptible variants, and the establishment of resistant variants following the treatment. To capture these effects, we construct an individual-based stochastic model for viral rebound (Fig.~\ref{Fig1}B). We specify a coarse-grained phenotypic model, where a viral strain of type $a$ is defined by a binary state vector $\vec \rho^a = [\rho^a_1,\dots,\rho^a_\ell]$, with $\ell$ entries for potentially escape-mediating epitope sites; the binary entry of the state vector at the epitope site $i$ represents the presence ($\rho^a_i=1$) or absence ($\rho^a_i=0$)  of a escape mediating mutation against a specified bNAb at this site of  variant $a$. We assume that a variant is resistant to a given antibody if at least one of the entries of its corresponding state vector is non-zero.

At each generation, a phenotypic variant $a$ can undergo one of three processes: birth, death and mutation to another type $b$, with rates $\beta_a$, $\delta_a$, and $\mu_{a \to b}$, respectively (Fig.~\ref{Fig1}B).  The net growth rate of  variant $a$  is its birth rate minus the death rate, $\gamma_a= \beta_a - \delta_a$ (Fig.~\ref{Fig1}C).  The total rate of events (birth and death) $\lambda = \beta_i + \delta_i$ modulates the amount of stochasticity in this birth-death process (Methods), which we assume to be constant across phenotypic variants.  The continuous limit for this birth-death process results in a stochastic evolutionary dynamics  for the sub-population of size $N_a$,
\EQA
{\small \frac{dN_a}{dt}= \begin{cases}
\text{absence of bNAb {\bf or} if $a$ is resistant:} \\
 N_a \left(f_a - \phi \right) +   \sum_{b}\left( N_b \mu_{{}_{b\to a}} - N_a \mu_{{}_{a\to b}} \right)\\ \quad +   \sqrt{N_{a} \lambda } \, \eta(t)     \\\\
\text{presence of bNAb {\bf and} if $a$ is susceptible:} \\
-r N_a +  \sum_{b}\left( N_b \mu_{{}_{b\to a}}- N_a \mu_{{}_{a\to b}} \right) +   \sqrt{N_{a} \lambda } \, \eta(t)  
\end{cases} }
\label{Langevin}
\EEA 
where  $\eta(t) $ is a Gaussian random variable with mean $\langle \eta(t)\rangle =0$ and correlation $\langle \eta(t) \eta(t')\rangle =\delta(t-t')$ (Methods). Here, $f_a$ denotes  the intrinsic  fitness of variant $a$ and its net growth rate $\gamma$ is mediated by a competitive pressure ${\small \phi =  \frac{1}{N_k} {\sum_{b} N_b f_b}}$ with the rest of the  population constrained by the carrying capacity $N_k$, such that $\gamma_a = f_a-\phi$.  In the presence of a bNAb, birth is effectively halted for susceptible variants and their death rate is set by the neutralization rate of the antibody, resulting in a net growth rate, $\gamma_\text{sus.} = - r$. At the carrying capacity, the competitive pressure is the mean population fitness  $\phi=\overline{f}$, making the net growth rate of the whole population zero (Fig.~\ref{Fig1}C). When susceptible variants are neutralized by a bNAb, the competitive pressure $\phi$ drops, and as a result, the resistant variants can rebound to carrying capacity, at growth rates near their intrinsic fitness.

To connect the birth-death model (eq.~\ref{Langevin}) with data, we should relate the simulation parameters of a birth-death process to  molecular observables. We have already made a connection between the birth and death rates of a variant  and its   intrinsic fitness in eq.~\ref{Langevin}.  In addition, the neutral diversity $\theta$ of a population at steady-state can  be expressed as $ \theta = {2 N_k \mu}/{ \lambda}$, where $\mu\approx 10^{-5} / \text{ day} $ is the per-nucleotide mutation rate, which we infer from intra-patient longitudinal  HIV sequence data~\cite{Zanini:2015gg}  (Methods). For consistency,  we set the  total rate of events $\lambda$ to be at least as large as the fastest process in the dynamics, which in this case is the growth rate of susceptible viruses $\gamma  \approx (3 \text{ days})^{-1}$, we choose $\lambda  = (0.5 \text{ days})^{-1}$ (Methods). Therefore, the key parameters of the birth-death model, i.e., $ \beta,\, \delta,\, \text{ and } N_k$ can be expressed in terms of the intrinsic fitness of the variants $f_a$ and the neutral diversity $\theta$, which we will infer from data.

\section{Results}
\subsection*{Population genetics of HIV escape from bNAbs} 
HIV escape from different bNAbs has been a subject of interest for vaccine and therapy design, and a number of escape variants against different bNAbs have been identified  in clinical trials or in infected individuals~\cite{lynchHIV1FitnessCost2015,Caskey:2015hm,scheidHIV1Antibody3BNC1172016,Caskey:2017el,bar-onSafetyAntiviralActivity2018}. This {\em in-vivo} data is often complemented with information from co-crystallized structures of bNAbs with the HIV envelope protein~\cite{Pancera:2017dw}, and {\em in-vitro} deep mutational scanning (DMS) experiments, in which the relative change in the growth rate of tens of thousands of viral mutants are measured in the presence of different bNAbs~\cite{Dingens:2017bk,Dingens:2019fd,Schommers:2020er}.  We identify escape mutations against each of the bNAbs in this study by using information from clinal trials,  the characterized binding sites, and the DMS assays (Methods); the list of escape mutations against each bNAb is given in Table~S1. 

\begin{figure*} [t!]
    \centering
	\includegraphics[width=0.85\textwidth]{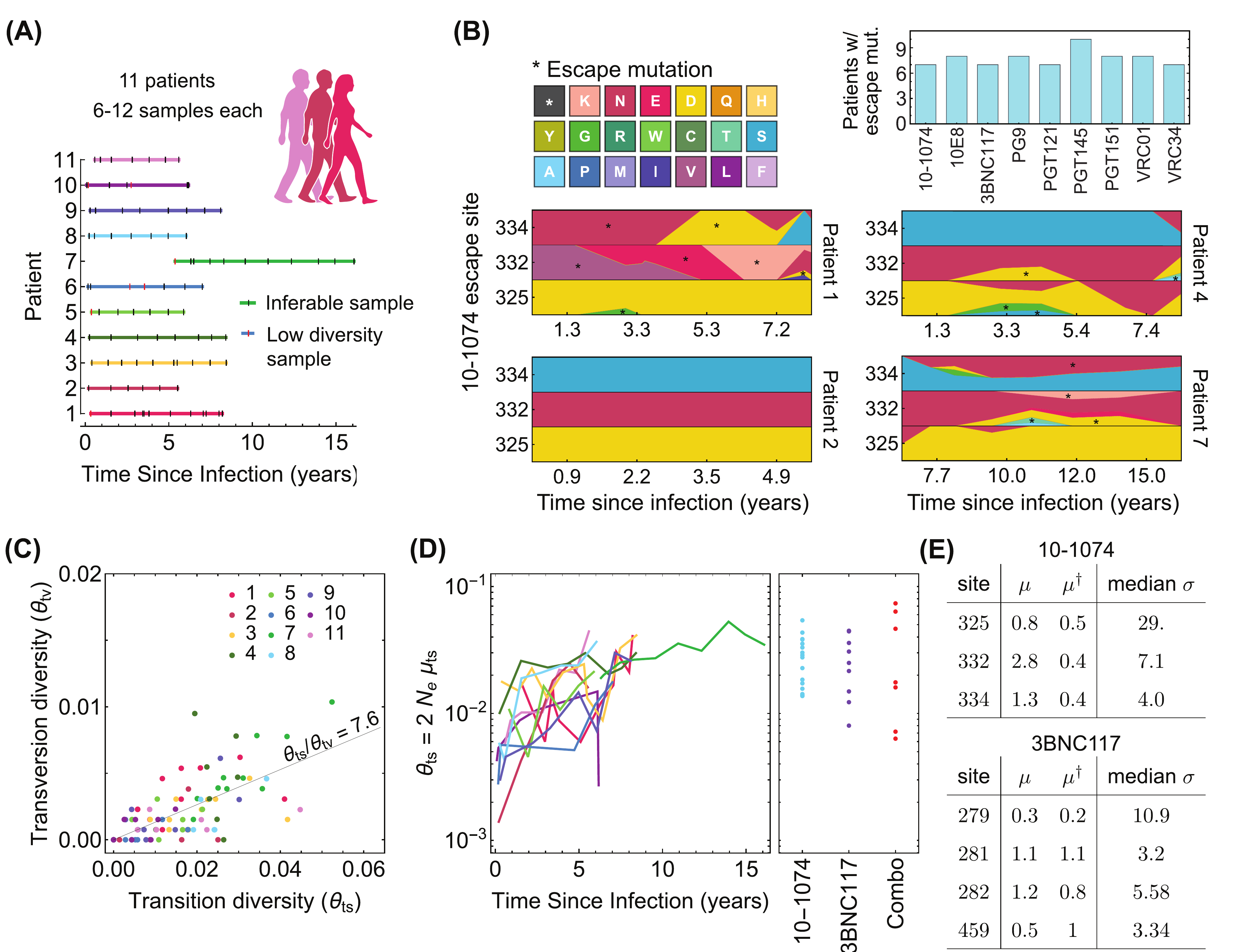}
    \caption{
    {\bf Statistics of the genetic data from bNAb-naive HIV patients.} 
{\bf (A)} 
	Statistics of the high throughput longitudinal data collected from  HIV populations in 11 ART-naive patients from ref.~\cite{Zanini:2015gg}, is shown. Some of the have low diversity (vertical red lines) and were not usable for our study. 
	Usable samples (vertical black lines) amount to 4-10 samples per patient, collected over 5 -10 years of infection.
{\bf (B)} 
	Lower panels show the relative frequencies (cube-root transformed for legibility) of different amino acids in four patients at the 3 escape sites against the 10-1074 bNAb, estimated from the polymorphism data  at the nucleotide level  in each patient over time. Despite 10-1074 being a broadly neutralizing antibody, mutations associated with escape (indicated by a $*$) are commonly observed in untreated patients. The upper right panel shows the number of individuals (out of the cohort of 11 bNAb-naive HIV patients) that carry mutations associated with escape against the indicated bNAbs. {\bf (C)}
	The nucleotide diversity associated with transversion $\theta_\tv= 2N_e \mu_\tv$ is shown against the transition diversity $\theta_\ts= 2N_e \mu_\ts$  for all patients (colors) and all time points.
	The covariance of these two diversity measurements yields an estimate for the transition/transversion ratio $\theta_\ts/\theta_\tv=7.6$.
{\bf (D)}
	Left: The transition diversity is shown to grow as a function of time since infection in all the 11 patients (colors according to (A)).
	Right: The neutral diversity of viral populations in patients (points) from the three different clinical trials~\cite{Caskey:2015hm,Caskey:2017el,bar-onSafetyAntiviralActivity2018} analyzed in  this study resemble the larger diversities of long-established viral populations in untreated patients.
{\bf (E)} The inferred forward and backward mutation rates ($\mu,\,\mu^\dagger$), relative to the transition rate, and the median selection strength $\sigma$ at each escape site against the two bNAbs (10-1074, and 3BNC117) from the trial data used in this study are shown. Compared to the 10-1074 bNAb, escape from the  3BNC117 bNAb appears to be less costly, and is associated with a smaller mutational target.}
   \label{Fig2}
\end{figure*}

The rise and establishment of an escape variant against a specific bNAb depend on three key factors, (i) {neutral} genetic diversity of the viral population, (ii) the mutational target size for escape from the bNAb and (iii) the intrinsic fitness associated with such mutations. Although viremia traces in clinical trials can be used to model the escape dynamics~\cite{Lu:2016id,Reeves:2020ca,Saha:2020fd,meijersPredictingVivoEscape2021}, they do not offer a comprehensive statistical description for HIV escape as they are limited by the number of  enrolled individuals. Alternatively, mutation and fitness characteristics of such escape-mediating variants can be inferred from a broader cohort of untreated and bNAb-naive patients. We  will infer these quantities from the large amount of high-throughput HIV sequence data from ref.~\cite{Zanini:2015gg} (see Fig.~\ref{Fig2}A,~B for details) and use them to parameterize the birth-death model (Fig.~\ref{Fig1}B).

\noindent{\bf Diversity of the viral population.} The neutral genetic diversity $\theta = 2N_e\mu/\lambda$ (i.e., the number of segregating alleles) determines the chance to observe a rare (e.g. resistant) mutation in a patient prior to treatment, and it modulates the strength of selection in an HIV population to escape a bNAb; here $N_e$ is the effective population size, $\mu$ is the per-nucleotide mutation rate, and $\lambda$ is the total number of events per virus in the birth-death process which determines the noise amplitude (Methods).  We use synonymous changes as a proxy for diversity  associated with the neutral variation in an HIV population at a given time point within a patient.  By developing a maximum-likelihood  approach based on the multiplicities of different synonymous variants, we can  accurately infer the neutral diversity of a population from the large survey of synonymous sites in the HIV genome (Methods and Fig.~S4).
Importantly, we infer the neutral diversity of transition $\theta_\text{ts}$ and transversion $\theta_\text{tv}$ mutations separately, and consistent with previous work~\cite{Feder:2016bc,Zanini:2017in}, find that transitions occur with a rate of about 8 times larger than transversions (Fig.~\ref{Fig2}C,~Fig.~S4B-E).

Our inference indicates that the neutral diversity grows over the course of an infection in untreated HIV patients from ref.~\cite{Zanini:2015gg} (Fig.~\ref{Fig2}D). The patients enrolled in the three bNAb trials~\cite{Caskey:2015hm,Caskey:2017el,bar-onSafetyAntiviralActivity2018} show a broad range of neutral diversity prior to bNAb therapy (Fig.~\ref{Fig2}D). In addition to the circulating viruses in a patient's sera, the viral reservoir, which consists of replication-competent HIV in latently infected cells or un-sampled tissue,  can also contribute to a bNAb escape in a patient. Evidence for the effect of reservoir is directly visible in trials as the failure of pre-trial sequencing to exclude
	patients who do not harbor escape variants.
We model the effect of the reservoir as augmenting the neutral diversity by a constant multiplicative factor $r_\text{resv.}$,
	so that patients with more diverse sera, representing usually longer infections, 
	are also expected to have correspondingly more diverse reservoir populations. By fitting the observed rebound data, we infer the reservoir factor $r_\text{resv.}\simeq 2.07$ (Methods, Fig.~S4). We use the  augmented genetic diversity  of HIV prior to the bNAb therapy in each trial  to generate the rebound time and the probability of HIV escape in patients. \\

\noindent{\bf Mutational target size for escape.} We define the mutational target size for escape from a bNAb as the number of trajectories that connect the susceptible codon to codons associated with escape variants, weighed by their probability of occurrences (Methods).  The connecting paths with only single nucleotide transitions or transversions dominate the escape and can be represented as connected graphs shown in Fig.~\ref{Fig1}D. To characterize the  target size of escape for each bNAb, we determine the forward mutation rate {$\mu \equiv \mu_{\text{sus.} \to \text{res.}}$ from the susceptible codons to the resistant (escape) codons, and the reverse mutation rate  $\mu^\dagger \equiv \mu_{\text{sus.} \leftarrow \text{res.}}$} back to the susceptible variant (Fig.~\ref{Fig1}D, Methods). The  mutational target sizes vary across  bNAbs,  with HIV escape being most restricted from 10E8 ($\mu/\mu_{\ts} = 1.8$) and most accessible in the presence of 10-1074 ($\mu/\mu_{\ts} = 4.9$); see Fig.~\ref{Fig4}C and Table~S1 for the list of mutational target size for escape against all bNAbs in this study.\\

\begin{figure*} [t!]
    \centering
	\includegraphics[width=0.82\textwidth]{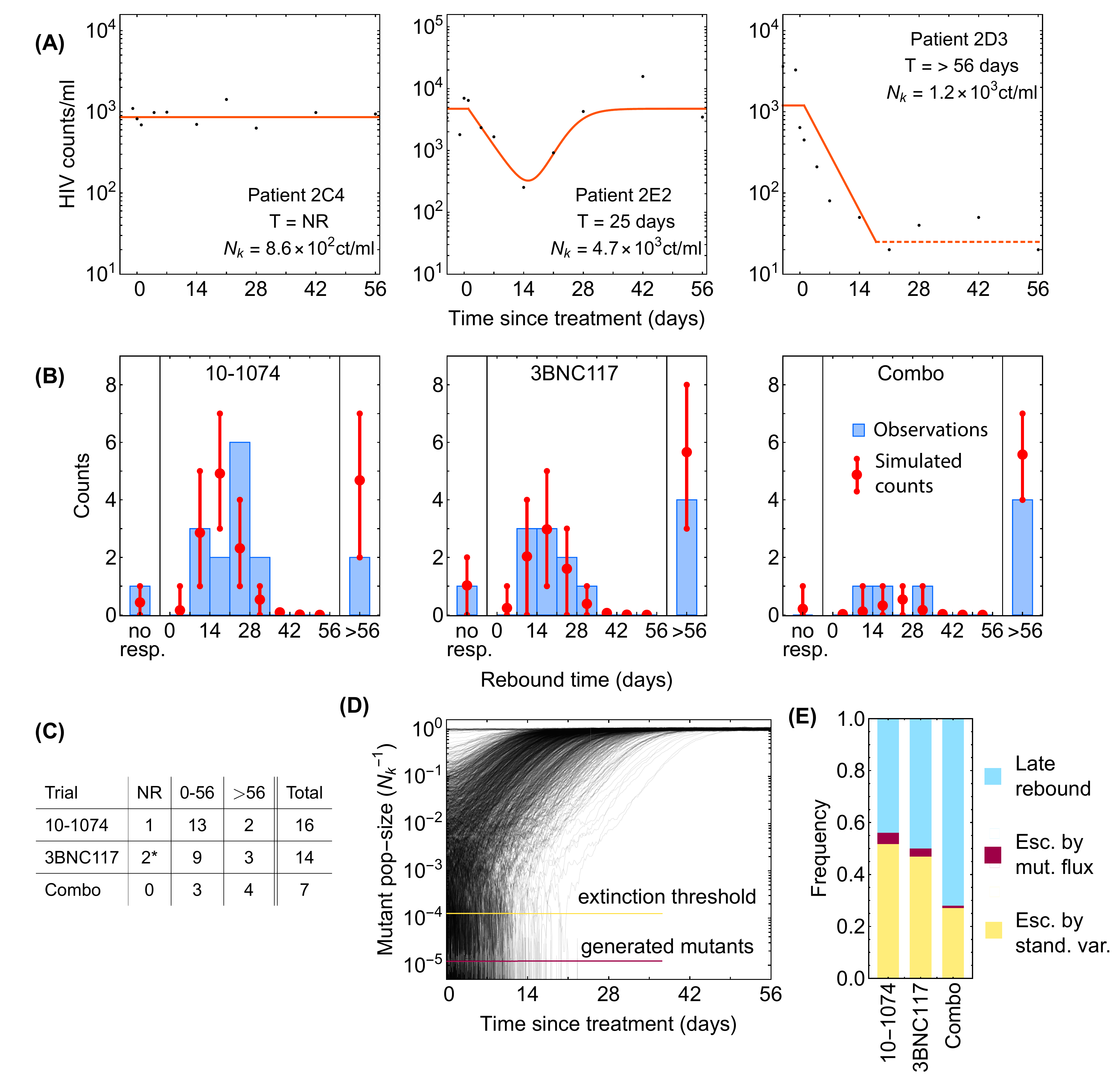}
    \caption{
    {\bf Statistics of  viral rebound  in clinical trials with bNAbs.}
    {\bf (A)} 
Panels show viremia  of three patients from  the 3BNC117 trial over time (black circles)  and the fitted model of the viral decay and rebound processes from eq.~\ref{eq:logisticpiecewise} (orange line). The viral rebound time $T$ and the fitted carrying capacity $N_k$ is shown in each panel. Shown are examples of a non-responder (NR; left),	 a rebound occurring during the trial window ($0<T<56$ days; center), and a late rebound ($T>56$ days; right).  Viremia traces from all patients and all trials are shown in Figs.~S1-S3.
    {\bf (B)} 
	We compare the distribution rebound times in patients from the three clinical trials with 10-1074~\cite{Caskey:2017el}, 3BNC117~\cite{Caskey:2015hm}, and the combination of the two bNAbs~\cite{bar-onSafetyAntiviralActivity2018} to the predictions from the simulations based on our evolutionary model (Fig.~\ref{Fig1}, and Methods). 
	The error bars show the inter decile range (0.1-0.9 quantiles) 
		generated by the simulations for the corresponding trial.
    {\bf (C)} The summary table shows the  number of patients for whom the infecting HIV population shows  no response (NR), rebound during the trial window $0<T<56$, and a late rebound  ($T>56$ days) in each trial.  Note that 3 patients were excluded from the 3BNC117 trial (*) because of insufficient dosage leading to weak viral response: $1\text{mg}/\text{kg}$ compared to the $3-30\text{mg}/\text{kg}$ in the other treatment groups.
    {\bf (D)} 
    	Plotted are $1,200$ trajectories of the mutant viral population simulated using our individual based model. 
    	Due to the individual birth-death events, fluctuations are larger when the population size is smaller.
	At a critical threshold, $x_\text{ext}$, fluctuations are large enough to lead to almost certain extinction in the existing viral population.
	The critical threshold (yellow line) is an order of magnitude larger than the post-treatment spontaneously-generated mutant fraction (red line).
    {\bf (E)}
    	The predicted fraction of escape events associated with post-treatment spontaneous mutations (red) and the pre-treatment standing variation (yellow) are shown for the three trials. The fraction of events associated with late rebound is indicated in blue. 	Because the spontaneously-generated mutant fraction is smaller than the extinction threshold, these mutations  contribute to less than 4\% of escape events (red), and  escape is likely primarily driven driven by standing variation (yellow), i.e., pre-existing escape variants.    }
   \label{Fig3}
\end{figure*}

\noindent{\bf Fitness effect of escape mutations.}  Since bNAbs target highly conserved regions of the virus, we expect HIV escape mutations to be intrinsically deleterious for the virus~\cite{Ferguson:2013kb,meijersPredictingVivoEscape2021}, and incur a fitness cost relative to pre-treatment baseline $f_0$. We assume that fitness cost associated with escape mutations are additive and background-independent so the fitness of a variant $a$ in the absence of bNAb follows, $f_a  =  f_0 - \sum_{i} \Delta_{i} \sigma_{i}^{a} $, where $ \Delta_{i}$ is the cost associated with the presence of a escape mutation at site $i$ (i.e., for $\sigma_{i}^{a}  =1$); see Fig.~\ref{Fig1}D. 

Interestingly, we observe the escape variants against different bNAbs to be circulating in the HIV populations from the cohort of ART- and bNAb-naive patients~\cite{Zanini:2015gg} (Fig.~\ref{Fig2}B). We use this data~\cite{Zanini:2015gg} and extract the multiplicity of susceptible and escape variants in HIV populations at each sampled time point  from a given patient. We use a single locus approximation under strong selection to represent the stationary distribution of the  underlying frequency of escape alleles $x$ in each patient from ref.~\cite{Zanini:2015gg}, 
	$P(x;\sigma, \theta,\theta^\dagger)\sim x^{-1+\theta} (1-x)^{-1+ \theta^\dagger} \exp[-\sigma x]$, 
	given the (scaled) fitness difference between the susceptible and the escape variants
	$\sigma= 2N_e (f_\text{sus}-f_\text{mut})$; see Methods.

Based on the statistics of escape and susceptible variants in all patients, we define a likelihood function that determines a Bayesian posterior for  selection $\sigma$ associated with escape at each site (Methods). We found that it is statistically more robust to infer the strength of selection relative to a reference diversity measure $\sigma/ \theta_\text{ts} = (f_\text{sus.}-f_\text{res.}) / \mu_\text{ts}$, for which we choose the transition rate (Methods). This approach generates unbiased selection estimates in simulations and is robust to effects of linkage and recombination (Methods and Figs.~S5,~S6). The inferred values of the scaled fitness costs $\sigma/ \theta_\text{ts}$ are shown for the   escape-mediating sites of the  trial bNAbs in Fig.~\ref{Fig2}E, and are reported in Table~S1.

\begin{figure*} [t!]
    \centering
	\includegraphics[width=0.9\textwidth]{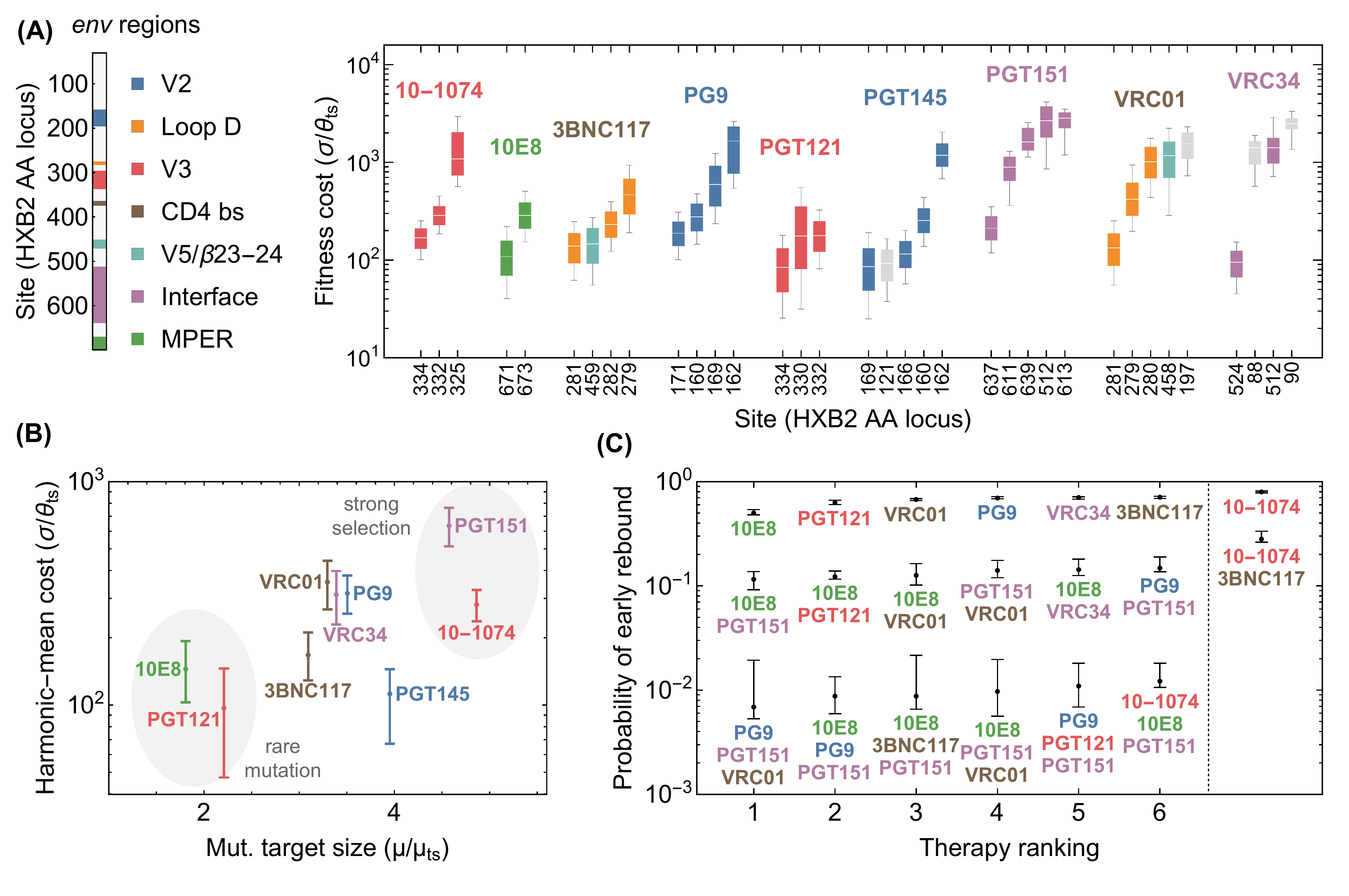}
    \caption{ {\bf Statistics of viral escape for optimal combination therapy with bNAbs.}
    {\bf  (A) } 
The posterior distribution for inferred selection strength on the escape-mediating sites associated with each of the 9 bNAbs in this study is shown (right); white line: median, box: 50\% around the median, bar: 80\%
around median.
Each escape site is color coded  by its location on the {\em env} gene (left) and each antibody by its associated epitope location.     {\bf (B)} The harmonic mean of the selection strength $\sigma$ associated with cost of escape (scaled by transition diversity $\theta_\ts$) is shown against the mutational target size for each bNAb; error bars indicate 50\% around the median.
For antibodies to be broadly-neutralizing, it is sufficient that viral escape from  them to be associated with a small mutational target size or a large fitness cost. The mutational target size is found to be weakly correlated with the average  cost of escape from a given bNAb. We identify two distinct strategies for antibody breadth---selection limited and mutational-target-size limited escape pathways each highlighted in gray.
    {\bf (C)}   
  bNAb therapies with 1, 2, and 3 antibodies are ranked based on the predicted probability of early  viral rebound, and in each case,  six therapies with  highest efficacies are shown; best ranked  therapy is associated with the lowest probability of early rebound; indicate 50\% around the median.
   Also for reference, the probability of early viral rebound two therapies from the trials in this study (10-1074 and 10-1074+3BNC117) are shown.
    }
   \label{Fig4}
\end{figure*}

\subsection*{Predicting the efficacy of bNAb therapy in clinical trials}
Monotherapy trials with 10-1074~\cite{Caskey:2017el} and with 3BNC117~\cite{Caskey:2015hm}, and the combination therapy with both of these antibodies in ~\cite{bar-onSafetyAntiviralActivity2018} have shown variable outcomes. In some patients bNAb therapy did not suppress the viral load, whereas in others suppression was efficient and no rebound was observed up to 56 days after infusion (end of  surveillance in these trials); see Fig.~\ref{Fig3}A for  examples of patients with different rebound times, Figs.~S1-S3 
 for the viremia traces in all patients, and Figs.~\ref{Fig3}B,~C for the distributions and the summary statistics of the rebound times in patients  in different trials. 

Although we infer a large intrinsic fitness cost for a virus to harbor an escape allele (Fig.~\ref{Fig2}E), these variants can emerge or already be present due to the large intra-patient diversity of HIV populations (Fig.~\ref{Fig2}C), or a larger mutational target size for these escape variants. Deep sequencing data in untreated (likely bNAb-naive) patients shows circulation of resistant variants against a panel of bNAbs in the majority of patients (Fig. \ref{Fig2}B). Our goal is to predict the efficacy of a bNAb trial, using the fitness effect and  the mutational target size for escape from a given bNAb, both of which we infer from the high-throughput HIV sequence data collected from bNAb-naive patients in ref.~\cite{Zanini:2015gg} (Figs.~\ref{Fig2}E, Table~S1). In addition, we modulate these measures with the patient-specific neutral diversity $\theta$ inferred from whole genome sequencing of HIV populations in each patient prior to bNAb therapy (Fig.~\ref{Fig2}D). These quantities parametrize the birth-death process for viral escape in a bNAb therapy (Fig.~\ref{Fig1}A), which we use to characterize the  distribution of rebound times in a given trial (Methods).

For both the  the 3BNC117 and the 10-1074 trials~\cite{Caskey:2015hm,Caskey:2017el}, we see an excellent agreement between our predictions of the rebound time distribution and data; see Fig.~\ref{Fig3}B, and Methods and Fig.~S7 for statistical accuracy of this comparison. By assuming an additive fitness effect for escape from 10-1074 and  3BNC117, we also accurately predict the distribution of rebound times  in the combination therapy~\cite{bar-onSafetyAntiviralActivity2018} (Fig.~\ref{Fig3}B). The agreement of our results with  data for combination therapy is consistent with the fact that the escape mediating sites from 10-1074 and 3BNC117 are spaced farther apart on the genome than 100bp, beyond which linkage disequilibrium diminishes due to  
frequent recombination in HIV~\cite{Zanini:2015gg}. Importantly, in all the trials, our evolutionary model accurately predicts the fraction of participants for whom  we should expect a late viral rebound (more than 56 days passed bNAb infusion)---the quantity that determines the efficacy of a treatment. 

Apart from the overall statistics of the rebound times, our stochastic model also enables us to characterize the relative contributions of the pre-treatment standing variation of the HIV population versus the spontaneous  mutations emerging during a trial to viral escape from a given bNAb. Given the large population size of HIV  and a high mutation rate ($\mu =10^{-5} \text{nt}$ per generation), spontaneous mutations generate a fraction $x^{(\mu)}$  of resistant variants during a trial, which we can express as, 
\begin{align}
x^{(\mu)}&=  \int_0^{56 \text{ wks}} (1-x(0))  e^{-r t } \mu \gamma dt \end{align}
In the best case scenario, there are no resistant virions prior to treatment i.e., $x(0)=0$. Since the neutralization rate $r$ and the growth rate $\gamma$ are comparable, this deterministic approach predicts that mutations can generate a resistant fraction of $x^{(\mu)} \approx 10^{-5}$ during a trial. 
However, stochastic effects from random birth and death events play an important role in the fate  and establishment of these resistant variants. The probability of extinction for a variant at frequency $x$ can be approximated as  $p(\text{extinct}) \approx 1- e^{- x/ x_\text{ext}}$ (Methods); here $x_\text{ext}= \frac{\mu_{ts}}{\gamma \theta_{ts}}\approx 10^{-4}$ and a variant with fraction $x$ that falls below this critical value is likely to go extinct (Fig.~\ref{Fig3}D). Since the total integrated mutational flux fraction during a trial is $x^{(\mu)} \sim 10^{-5}$, mutational flux rarely decides the outcome of patient treatment. Indeed we infer that spontaneous mutations contribute to less than 4\% of escape events in all the three trials and  escape is primarily attributed to the standing variation prior from the serum or the reservoirs prior to therapy (Fig.~\ref{Fig3}E). A similar conclusion was previously drawn based on a mechanistic model of escape in VRC01 therapy trials~\cite{Saha:2020fd}.

\section{Devising optimal bNAb therapy cocktails}
Clinical trials with bNAbs have been instrumental in demonstrating the potential role of bNAbs  as therapy agents and in measuring the efficacy of each bNAb to suppress HIV. Still, these clinical trials {can only test a small fraction of the potential therapies that can be devised. It is therefore important that trials test therapies that have been optimized based on surrogate estimates of treatment efficacy.}  The accuracy of our predictions  for the rebound time of a HIV population subject to  bNAb therapy suggests a promising approach to the rational design of therapies based on genetic data of HIV populations collected from bNAb-naive patients. 
	
	Here, we use genetic data to infer the efficacy of therapies with bNAbs, for which clinal trials  are not yet performed. To do so, we first need to identify the routes of HIV escape from these  bNAbs. We  use deep mutational scanning data on HIV subject to 9 different bNAbs from ref.~\cite{Dingens:2019fd} together with information from literature  to identify the escape mediating variants from each of these bNAbs (Methods and Table~S1). We then determine the mutational target size and the fitness effect of these escape variants using high-throughput sequences of HIV in bNAb-naive patients from ref.~\cite{Zanini:2015gg}; these inferred values are reported in Fig.~\ref{Fig4}A,~B and Table~S1. Using the inferred fitness and mutational parameters and by setting the pre-treatment neutral diversity $\theta$ to  be comparable to that of the patients in the previous three trials with 10-1074 and 3BNC117 (Fig.~\ref{Fig2}C), we simulate treatment outcomes for these 9 bNAbs and their 2-fold and 3-fold combinations. 
	
	Interestingly, we infer that escape from mono-therapies is almost certain and a  combination of at least 3 antibodies is necessary to limit the probability of early rebound to below 1\% (Fig.~\ref{Fig4}C).  When considering all nine antibodies together, interesting  patterns emerge. 
	We find that the mutational target size and the fitness cost of escape, estimated as the harmonic-mean selection cost of individual sites, obey a roughly linear relationship (Fig.~\ref{Fig4}B). As all these bNAbs have similar overall breadth (i.e., they neutralize over 70\% of panel strains), this result suggests that for an antibody to be broad, its escape mediating variants should either be rare  (i.e, small mutational target size) or  intrinsically costly  (i.e., incurring a high fitness cost), but it is not necessary to satisfy both of these requirements. For instance, we find that 10E8 has a relatively weak selection but has a small escape target size, while PGT151 has a larger escape target size but makes up for it by having mutants with unavoidably high fitness cost. 
	
The fitness-limited versus the mutation-limited  strategies have different implications for the design of combination cocktails. The small mutational target size of 10E8 makes it the best candidate antibody for mono-therapy among the antibodies we consider because because the escape variants against this antibody are less likely to circulate in a patient's serum prior to treatment. However, in combination, 10E8 appears less often in top ranked therapies than PGT151. PGT151 is unremarkable on its own because of a relatively large target size, but the high cost of escape makes it especially promising in combination therapies. Overall, fitness-limited bNAbs like PGT151 are more effective against high diversity viral populations,  while mutation-limited bNAbs such as 10E8 are more effective against low diversity viral populations. Indeed, the best ranked therapy, namely the combination of PG9, PGT151, and VRC01, combines antibodies that target different regions of the virus and also have both types of fitness- and  mutation-limited strategies for coverage against the full variability of viral diversities found in pre-treatment individuals (Fig.~\ref{Fig4}C) participating in the clinical trials.

\section{Discussion}
HIV therapy with passive bNAb infusion has become a promising alternative to anti-retroviral drugs for suppressing and preventing the disease in patients without a need for daily administration. The current obstacle is the frequent escape of the virus seen in  mono- and even combination bNAb therapy trials~\cite{Caskey:2015hm,Bar:2016hg,Caskey:2017el,bar-onSafetyAntiviralActivity2018}. The key is to identify bNAb cocktails that can target multiple vulnerable regions on the virus in order to reduce the likelihood for the rise of resistant variants with escape-mediating mutations  in all of these regions.  Identifying an optimal bNAb cocktail can be a combinatorially difficult problem, and designing patient trials for all the potential combinations is a costly pursuit.

Here, we have proposed a computational approach to predict the efficacy of a bNAb therapy trial based on population genetics of HIV escape, which we parametrize using high-throughput HIV sequence data collected from a separate cohort of bNAb-naive patients~\cite{Zanini:2015gg}. Specifically,  we infer the mutational target size for escape and the fitness cost associated with escape-mediating mutations in the absence of a given bNAb. These quantities together with the neutral diversity of HIV within a patient parametrize our stochastic model for HIV dynamics subject to bNAb infusion, based on which we can accurately predict  the distribution of rebound times for HIV in therapy trials with 10-1074, 3BNC117 and their combination.  Consistent with previous work on VRC01~\cite{Saha:2020fd}, we found that viral rebounds in  bNAb trials are primarily mediated by the escape variants present either in the patients' sera or their latent reservoirs prior to treatment, and that the escape is not {likely to be driven} by the emergence of spontaneous mutations that establish during the therapy.

One key measure of  success for a bNAb trial is the suppression of early viral rebound. Our model can accurately predict the rebound times of HIV subject to three distinct therapies~\cite{Caskey:2015hm,Caskey:2017el,bar-onSafetyAntiviralActivity2018},  based on the fitness and the mutational characteristics of escape variants inferred from high-throughput HIV sequence data. This approach enables us to characterize routes of HIV escape  from other bNAbs, for which therapy trials are not available,   and to design optimal therapies. We used deep mutational scanning data~\cite{Dingens:2019fd} to identify escape-mediating variants against 9 different bNAbs for HIV. Our genetic analysis shows that bNAbs gain breadth and limit viral escape either due to their small mutational target size for escape or because of the large intrinsic fitness cost  incurred by  escape mutations. bNAbs with mutation-limited strategy are more effective at preventing escape in patients with low viral genetic diversity, while bNAbs selection-limited strategy with more effective at high viral diversity. To suppress the chance of viral rebound  to below  $1\%$, we show that a combo-therapy with 3 bNAbs with a mixture of mutation- and selection-limited strategies that target different regions of the viral envelope is necessary. Such combination can counter the full variation of viral diversity observed in patients. We found that  PG9, PG151, and VRC01, which respectively target V2 loop, Interface, and CD4 binding site of HIV envelope, form an optimal combination for a 3-bNAb therapy to limit HIV-1 escape in patients infected with clade B of the virus.

We rest our analysis primarily on the predictive power of the  observed variant frequencies in the untreated patients. Our model weighs these frequencies with respect to the viral diversity in a mathematically and biologically consistent way.  However, we ignore the dynamics of antibody concentration and IC50 neutralization during treatments, the details of T-cell dynamics during infection~\cite{Perelson:2002kw}, and also the evolutionary features of the genetic data, such as epistasis between loci~\cite{Bonhoeffer:2004cf,Zhang:2020bs}, genetic linkage~\cite{Zanini:2015gg}, and codon usage bias~\cite{Meintjes:2005ji}. The statistical fidelity of our model to the observed variant frequencies implies that many of the mechanistic details are at best secondary to the coarse-grained features  of  viral escape, and specifically the distribution of the rebound times. Nonetheless, our approach falls short of predicting the detailed characteristics of viremia traces in patients, especially at very short or very long times,  during which the dynamics of T-cell response or the decay of bNAbs could play a role~\cite{Lu:2016id,Reeves:2020ca,Saha:2020fd}.

Our predictions are limited by our ability to identify the escape variants for each bNAb, either based on  trial data, patient surveillance, or {\em in-vitro} assays such as DMS experiments. DMS data for viral escape~\cite{Dingens:2017bk,Dingens:2019fd,Schommers:2020er} are generally of high quality but consider only a single HIV genetic background (e.g. in \cite{Dingens:2017bk} this background is BF520.W14M.C2), whereas clinical data will have diverse baseline viruses between and within individuals. In addition, DMS data can be noisy for variants that grow very poorly in the absence of a bNAb, since  growth without antibodies is the first stage  of these experiments, followed by   growth  in the presence of  bNAbs. This is likely to be the reason that the DMS experiments  fail to capture a clear escape signal against bNAbs such as VRC01 or 3BNC117 that target the CD4 binding site of HIV, where mutations can be extremely deleterious  in the absence of the bNAb. As such, we continue to need more data and more powerful statistical techniques for calling escape variants, using {\it in-vivo} approaches. For example, passive infusion of bNAb  in humanized mouse models~\cite{Gruell:2017}, which can, at least partially, reproduce the natural diversity of a human infection would be valuable. These efforts would take us a step closer to rational design for  bNAb therapy and a more model-guided clinical trials.

Our approach showcases that, when feasible,  combining high-throughput genetic data with ecological and population genetics models can have surprisingly broad  applicability, and their interpretability can shed light  into the complex dynamics of these populations.  Application of similar methods to therapy design to curb the escape of cancer tumors against immune- or chemo-therapy, the resistance in bacteria against antibiotics, or the escape of  seasonal influenza against vaccination is a promising avenue for future work. However, we expect that more  sophisticated methods for inferring fitness from evolutionary trajectories  may be necessary to capture the dynamical response of these populations.

\section*{Acknowledgements}
This work has been supported by the NSF CAREER award (grant No:~2045054), DFG grant (SFB1310) for Predictability in Evolution, and the MPRG funding through the Max Planck Society.

\bibliographystyle{plos2}

\end{document}


\noindent{\Large \bf Supplementary Information}\\\\
{\bf Design of an optimal combination therapy with broadly neutralizing antibodies to suppress HIV-1}\vspace{0.1cm}\\
Colin LaMont, Jakub Otwinowski, Kanika Vanshylla, Henning Gruell, Florian Klein, Armita Nourmohammad\\

\renewcommand{\theequation}{S\arabic{equation}} 
\renewcommand{\thetable}{S\arabic{table}}  
\renewcommand{\thefigure}{S\arabic{figure}}

\tableofcontents

\section*{Data and code accessibility:}
The code for the algorithms used in this work and the data are available on GitHub at \url{https://github.com/StatPhysBio/HIVTreatmentOptimization} and in the Julia package  \url{https://github.com/StatPhysBio/EscapeSimulator}.

\section{Description of molecular data}
\paragraph{Data from bNAb trials} In this study we considered three clinical trials for passive therapy with bNAbs:
\begin{itemize}
\item  Monoclonal therapy with 3BNC117  bNAb~\cite{Caskey:2015hm} with 16 patients enrolled, 13 of whom were off   anti-retroviral therapy (ART).
\item Monoclonal therapy with 10-1074 bNAb~\cite{Caskey:2017el}, with 19 patients enrolled, 16 of whom were off ART.
\item Combination therapy with 10-1074 + 3BNC117 bNAb~\cite{bar-onSafetyAntiviralActivity2018}, with 7 patients off ART.
\end{itemize}
All sequenced patients across all trials were infected with  distinct HIV-1 clade B viral strains. We limited our analyses to those patients not on ART at the time of treatment initiation. In these studies, the injected bNAb level falls off over time  within patients and therefore, we only considered dynamics within an 8 week window since infusion. This assures that rebound is not confounded by a drop in bNAb below sensitive-strain neutralizing levels of $\text{IC}_{50_S} < 2 \mu \text{g/m}$.

We used single-genome sequence data of {\em env} collected from all patients in each trial to characterize the  diversity of HIV population within each patient shown in Fig.~2 available from European Nucleotide Archive (Accession no: PRJEB9618). The patient sequence data for each trial is available through ref.~\cite{Caskey:2015hm} GeneBank PopSet: 1036347437, 
ref.~\cite{Caskey:2017el} GenBank accession numbers KY323724.1 - KY324834.1, and ref.~\cite{bar-onSafetyAntiviralActivity2018}, GenBank accession numbers MH632763 - MH633255.\\

\paragraph{Longitudinal HIV sequence data from untreated patients} Single nucleotide polymorphism (SNP) data was obtained from ref.~\cite{Zanini:2015gg} and aligned to the HXB2 reference using  HIV-align tool~\cite{gaschenRetrievalOntheflyAlignment2001}. The dataset includes 11 patients observed for 5-8 years of infection, with HIV sequence data sampled over 6-12 time points per patient (Fig.~2).

Patient 4 and patient 7 were excluded from the original analysis done in ref.~\cite{Zanini:2015gg} because of suspected superinfection and failure to amplify early samples, respectively. These patients were included in our analysis, since (i) super-infection poses no additional difficulties for our tree-free procedure, and (ii) only time points with measurable viral diversity entered into our selection likelihood, which automatically limits our analysis to samples with   high quality sequences.

All patients were infected with  clade B of HIV, except for patient 6 (clade C), and patient 1 (clade 01\_AE).  We assessed the robustness of our inference to exclusion of these patients from our analysis in Section~\ref{sec:robustness}. Overall, our inference was not strongly affected by this choice (Fig.~\ref{Fig:S5}), and therefore, we included these patients in our main analyses to enhance the statistics with larger data.

For our analysis we considered only data reported in single nucleotide polymorphism (SNP) counts. The number of raw SNP counts are the result of amplification and must be converted into estimates for the number of pre-amplification template fragments. { Zanini et. al.} reported that on average about $10^2$ templates of amplicon were  associated with the fragments of the envelope ({\em env}) protein~\cite{Zanini:2015gg}.
We converted raw SNP counts into templates by normalizing to $120$ counts and rounding the resulting number to the nearest integer.

\section{Identifying escape-mediating variants against bNAbs}
The starting point for our analysis of HIV response to a given a bNAb  is the description of the escape-mediating amino acids in the HIV {\em env} protein. We use a combination of methods to identify the escape variants for a given bNAb. First, we use deep mutational scanning (DMS) data  of HIV-1 in the presence of a bNAb from ref.~\cite{Dingens:2019fd} to identify these mutations. 
These DMS experiments  have created libraries of all single mutations from a given genomic background of HIV and tested the fitness of these variants (i.e., growth on T-cell culture)  in the absence and presence of 9 different bNAbs, including the 10-1074 and 3BNC117~\cite{Dingens:2019fd}.

Escape variants in DMS data are identified as those which are strongly selected for only in the presence of a bNAb. Specifically, we identify escape variants as those which show  3-logs-change in their frequency in the presence  versus  absence of  a bNAb. DMS data reflects {\em in-vitro} escape in cell culture. However, some of these variants may not be viable { in vivo}. To identify the reasonable candidates of escape {in vivo}, we limit our set to the variants  that are also observed  in  the circulating viral strains of untreated HIV-1 patients from ref.~\cite{Zanini:2015gg}. It should be noted that since we use HIV sequence data from ref.~\cite{Zanini:2015gg} to  infer selection on escape mediating variants in the absence of a bNAb, the candidates of escape that are not observed in the dataset~\cite{Zanini:2015gg}  would be inferred to be strongly deleterious, and hence, unlikely to contribute to our predictions of viral rebound.   

Our analysis of DMS data results in  a set of escape mediating amino acids for 10-1074 that is consistent the escape variants that emerge in response to the bNAb trial~\cite{Caskey:2017el} (Table~\ref{tab:full_antibody_data}). However, the DMS data is very noisy for bNAbs that target CD4 binding site of HIV, i.e.,  3BNC117 and VRC01~\cite{Dingens:2019fd}. 
One reason for this observation may be  that  the CD4 binding site is crucial for the entry of HIV  to the host's T-cells and mutations in this region are highly deleterious. As a results,  only a small number of variants with mutations in this region can survive in the absence of a bNAb in a DMS experiment. Growth in the absence of a bNAb is the first step in the DMS experiments, which is then followed by exposure of the replicated variants to a bNAb. Therefore, a low multiplicity of variants in the absence of a bNAb could  result in a noisy pattern of growth of the small subpopulation in the next stage of the experiment, in which growth is subject to a CD4-targeting bNAb.

For the CD4 binding site antibodies 3BNC117 and VRC01, we used additional data to call the escape variants. For 3BNC117 we used a combination of trial-patient sequences~\cite{Caskey:2015hm,scheidHIV1Antibody3BNC1172016}, 
i.e. post-treatment enrichment, along with contact site information compiled in the  crystallographic studies to narrow down candidate sites~\cite{zhouStructuralRepertoireHIV1neutralizing2015, labrancheHIV1EnvelopeGlycan2018}.
For VRC01 we assumed a similar escape pattern to 3BNC117 but included sites known from other studies \cite{lynchHIV1FitnessCost2015} and the clear DMS signal at HXB2 site 197.
The sites we called were similar to those identified using humanized-mouse models of HIV infection~\cite{horwitzHIV1SuppressionDurable2013a}, although more complex mutational patterns were seen in the soft-randomization scanning of~\cite{otsukaDiversePathwaysEscape2018}.
Although the complete list of escape substitutions are unknown and background-dependent~\cite{otsukaDiversePathwaysEscape2018}, the escape profiles which are most important are those that are most likely to be seen consistently in data and to be correctly identified.
The list of substitutions are shown in Table~\ref{tab:full_antibody_data}.

\section{Statistics and dynamics of viral rebound}

\subsection{Inference of growth parameters from dynamics of viremia}

\begin{figure*} [h!]
    \centering
	\includegraphics[width=1\textwidth]{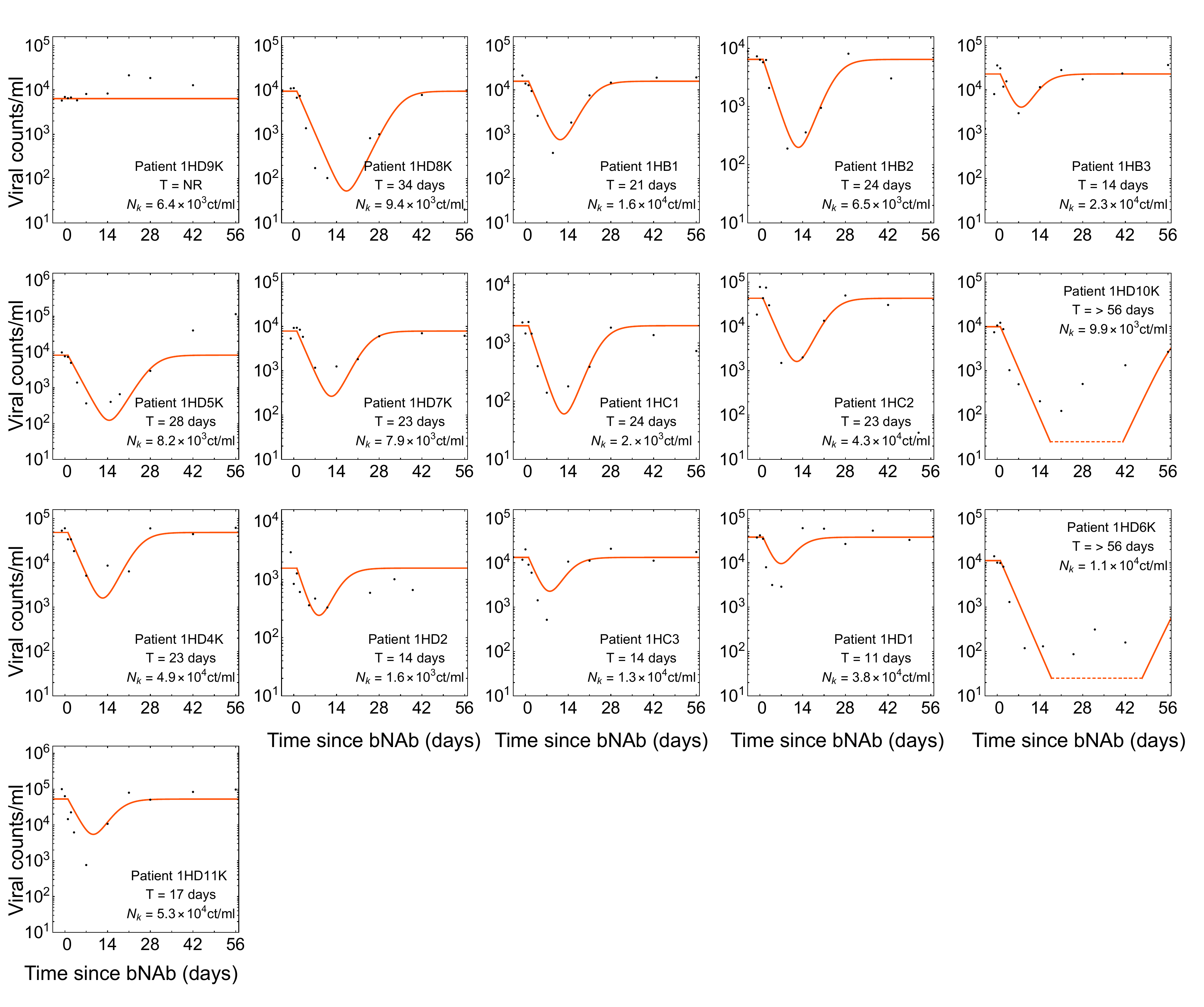}
    \caption{
    {\bf Dynamics of   viremia  in patients from the 10-1074 trial.} Panels shows the  measured viremia over time (black dots)  and the fitted deterministic dynamics from eq.~\ref{eq:logisticpiecewiseSI} (red line) for all the patients in the 10-1074 trial~\cite{Caskey:2017el}. Patient-specific fitted carrying capacity $N_k$ and the estimated rebound times are reported in each panel. A common decay rate $r=0.36 \text{ day}^{-1}$ is fitted to all patient data in this trial, and growth rate is set to $0.33 \text{ day}^{-1}$.
}
   \label{Fig:S1}
\end{figure*}

\begin{figure*} [t!]
    \centering
	\includegraphics[width=1\textwidth]{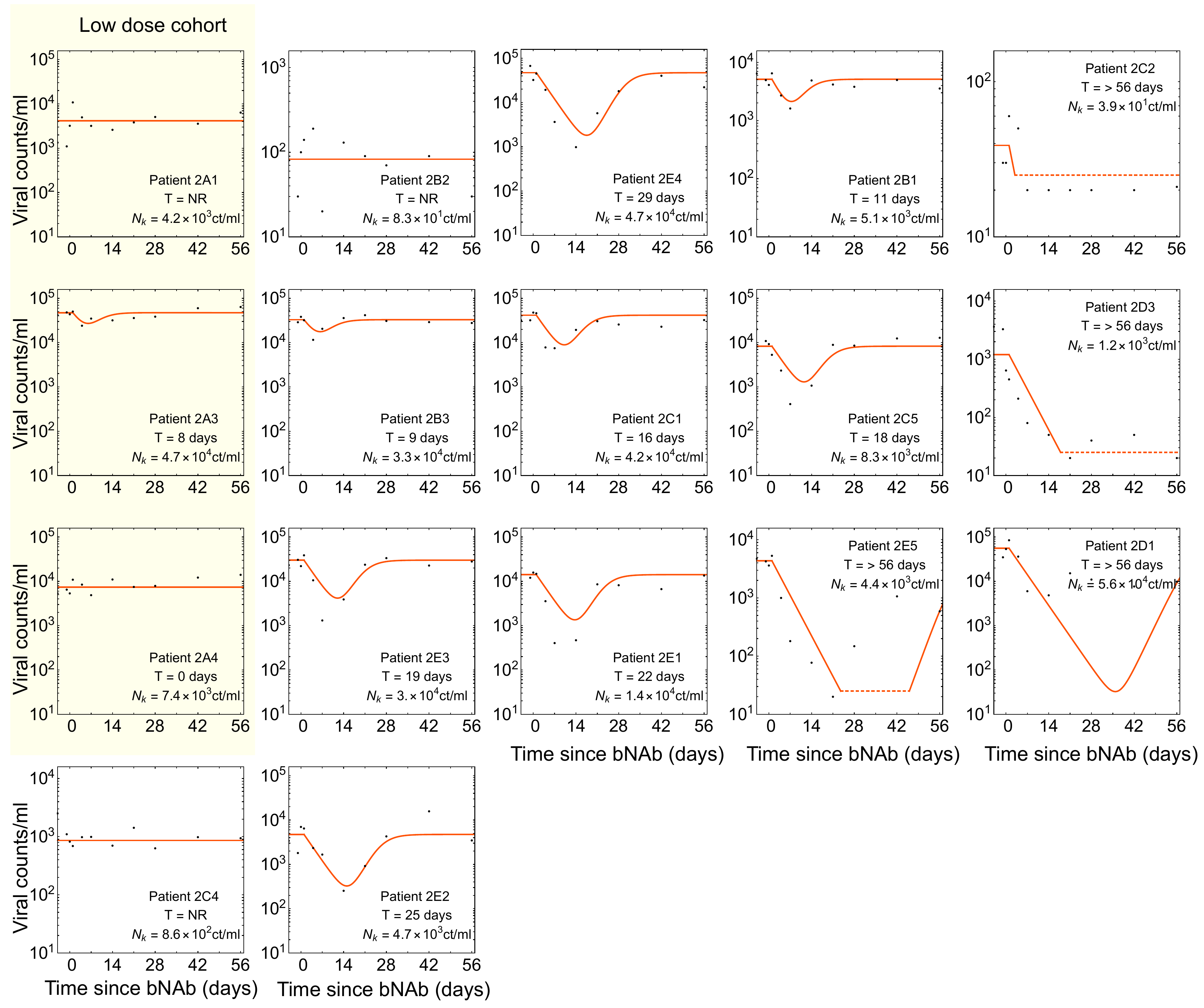}
    \caption{
    {\bf Dynamics of   viremia  in patients from the  3BNC117 trial.}  Similar to Fig.~\ref{Fig:S1} but for the 3BNC117 trial \cite{Caskey:2015hm}. The cohort treated with a low dosage of bNAb  ($1 \text{ mg}/\text{kg}$ as opposed to $3 - 30 \text{ mg}/\text{kg}$) is shown in yellow;  these patients exhibited a very weak response to there treatment and were excluded from our analysis. A common decay rate $r= 0.23 \text{ day}^{-1}$ is fitted to all patient data  in this trial, and growth rate is set to $0.33 \text{ day}^{-1}$.}
   \label{Fig:S2}
\end{figure*}

\begin{figure*} [t!]
    \centering
	\includegraphics[width=1\textwidth]{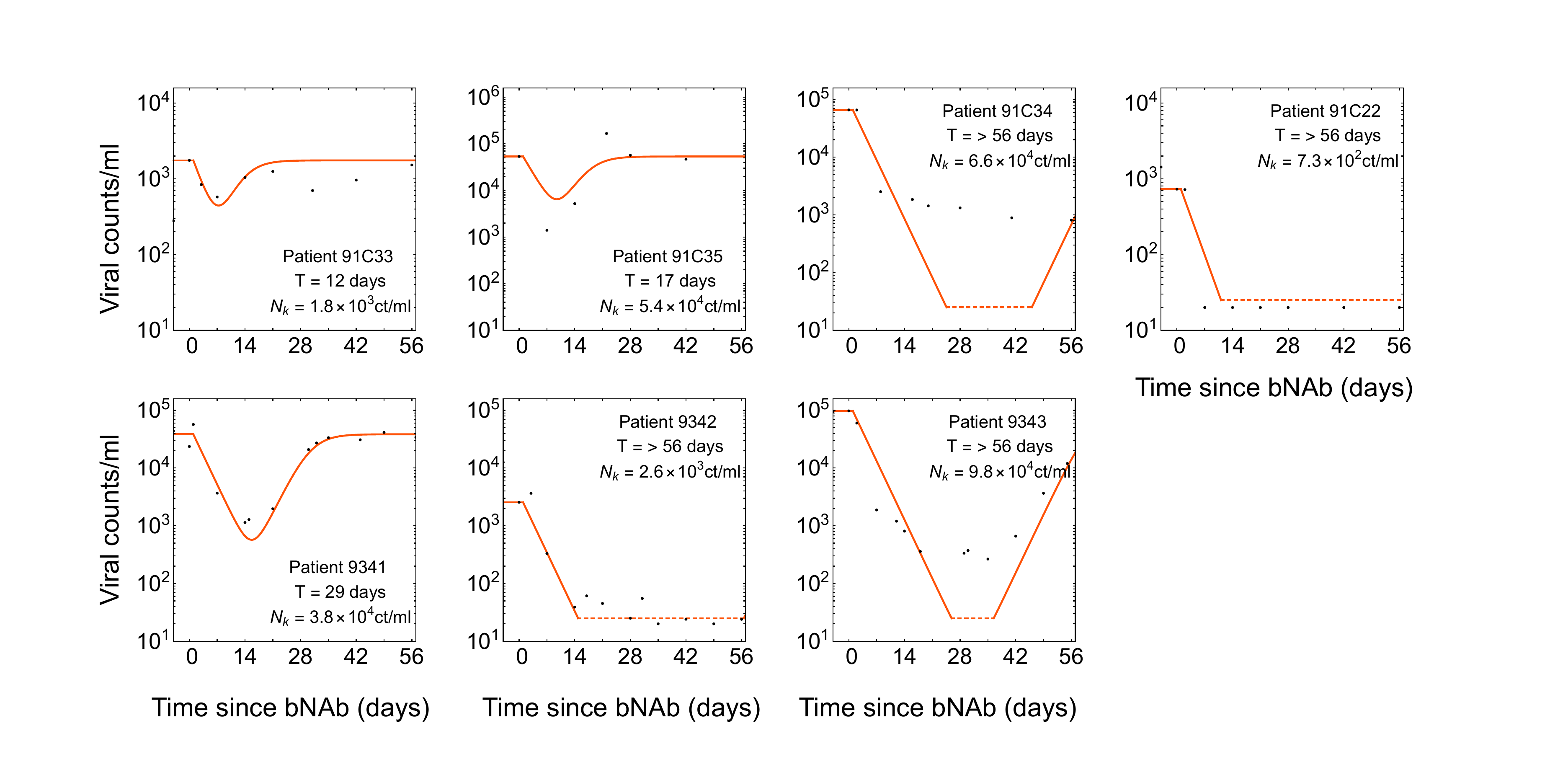}
    \caption{
    {\bf Fits to raw viremia data in combination trial} 
   Similar to Fig.~\ref{Fig:S1} but for the combination therapy with 10-1074 and  3BNC117~\cite{bar-onSafetyAntiviralActivity2018}.  A common decay rate  $r=0.33 \text{ day}^{-1}$ is fitted to all patient data  in this trial, and growth rate is set to $0.33 \text{ day}^{-1}$. 
}
  \label{Fig:S3}
\end{figure*}

The concentration of viral RNA copies in blood serum is a delayed reflection of the total viral population size $N(t)$, containing a resistant and susceptible subpopulations, with respective sizes $N_r(t)$ and $N_s(t)$. 
After infusion of bNAbs in a patient, the susceptible sub-population decays due to neutralization by bNAbs and the resistant sub-population grows and approaches the carrying capacity $N_k$, with the dynamics,  
		\begin{align}
\nonumber \frac{d N_r }{d t} &= \gamma N_r (1- N_r/N_k)\\
\nonumber \frac{d N_s }{d t} &= -r N_s\\
\label{eq:rebounddynamics}
\end{align}
Here, $\gamma$ is the growth rate of the resistant population, and $r$ is the neutralization rate impacting the susceptible subpopulation. By setting the initial condition for fraction of resistant subpopulation prior to treatment  (at time $t=0$) $x=  N_r(0)/ (N_r(0) + N_s (0))$, we can characterize the evolution of the total viremia in a patient. This dynamics is governed  by the combined processes of neutralization by the infused bNAb and the viral rebound (Fig.~1A), which entails,
\begin{align}
\label{eq:logisticpiecewiseSI}
N(t) = 
\begin{cases}
N_k 
	& t \leq 0 \\
(1-x) N_k e^{-r t} + \frac{N_k}{1+ \frac{1-x}{x}e^{- \gamma t}}
	& t>0
\end{cases}
\end{align}
We use eq.~\ref{eq:logisticpiecewiseSI} to define the evolution of blood concentration of viral RNA sequences which is observed indirectly via noisy viremia measurement data from refs.~\cite{Caskey:2015hm,Caskey:2017el,bar-onSafetyAntiviralActivity2018}.
To connect the data with the simple model of viral dynamics  in eq.~\ref{eq:logisticpiecewiseSI}, we  fit the initial frequency of resistant mutants $x$ for each patient separately, and fit a global estimate  for  the decay rate of susceptible variants $r$ shared across all patients in a trial, using a joint maximum-likelihood procedure. In addition, we fix the  growth rate $\gamma$ to 1/3 days, corresponding to a doubling time of approximately 2 days \cite{garciaDynamicsViralLoad1999}.
 Our analyses indicate that the initial viremia decline lagged treatment by about 1 day (Figs.~\ref{Fig:S1}-\ref{Fig:S3}),  consistent with previous findings~\cite{ioannidisDynamicsHIV1Viral2000}), and  therefore, we included a 1-day lag between the fitted viremia response model and the treatment.

The number of viral RNA copies in a blood sample is subject to count fluctuations with respect to the true number of circulating virions in a given volume of the blood. We use a Poisson sampling model to define a likelihood for our model of viral population. The likelihood of observing $k=n_p(t)$ viral counts in a sample collected from patient $p$ at time $t$ is given by a Poisson distribution,
\begin{align}
p(k = n_p(t)| \eta = N_p(t)) = e^{-\eta} \frac{\eta^k}{k!} 
\label{eq.poissCountSI}
\end{align}
with rate parameter set by the model value of the viral multiplicity $N_p(t)$ (eq.~\ref{eq:logisticpiecewiseSI}). We use the Poisson likelihood in  eq.~\ref{eq.poissCountSI} to characterize an error model to fit the parameters of the viral dynamics in eq.~\ref{eq:logisticpiecewiseSI}. However, since the mean and variance of the Poisson distribution are related, combining data with different mean values $N_p(t)$ at different times and from different patients can cause inconsistencies in evaluations of errors in our fits.  To overcome this problem, we use a variance stabilizing transformation~\cite{mccullaghGeneralizedLinearModels2019} and define a change in variable $\hat n_p(t) = \sqrt{ n_p(t )}$. This transformed variable has a constant variance, and in the limit of large-sample size, it is Gaussian distributed with a mean and avariance given by, $\hat n_p(t)\sim \mathcal{N} ( \sqrt{\lambda},1/4)$. The constant variance of the transformed variable enables us to  combine data from all patients and time points, irrespective of the sample's viral loads, and fit the model parameters $(r,x_p)$ using  (non-linear) least-squares fitting of the function
\begin{align}
R(r, \{x_{p}\}) = \sum_{p:\text{ patients},t} \left( \sqrt{N_p(t|r,x_p)} - \sqrt{n_p(t)} \right)^2
\label{eq:lse}
\end{align}
Here,  $N_p(t|r,x_p)$ is the model estimate of  viremia in patient $p$ at time $t$ (eq.~\ref{eq:logisticpiecewiseSI}), given the  pre-treatment fraction of resistant variants $x_p$, and the decay rate $r$.

Note that the viremia measurements have a minimum sensitivity threshold of 20 RNA copies per ml. We treat the data points below the threshold of detection as missing data and if $n_p(t)$ is below the threshold of detection we impute $ n_p(t) = \text{min}(20, N_t)$.  

The fitted viremia curves for patients enrolled in the three bNAb trials under consideration are shown in Figs.~\ref{Fig:S1}-\ref{Fig:S3}, and the respective decay rates $r$ for each experiment are,
\begin{align}
\label{tab:decay_rates}
\begin{tabular}{r|lll|l}
trial & 10-1074 & 3BNC117 & Combination & Avg\\
\hline
fitted $r$ ($\text{days}^{-1}$) & 0.36 & 0.23 & 0.33 & 0.31
\end{tabular}.
\end{align}

\subsection{Individual-based model for viral population dynamics}
To encode for different viral variants,  we specify a coarse-grained phenotypic model, where a viral strain of type $a$ is defined by a binary state vector $\vec \rho^a = [\rho^a_1,\dots,\rho^a_\ell]$, with $\ell$ entries for potentially escape-mediating epitope sites; the binary entry of the state vector at the epitope site $i$ represents the presence ($\rho^a_i=1$) or absence ($\rho^a_i=0$)  of a escape mediating mutation against a specified bNAb at  site $i$ of  variant $a$. We assume that a variant is resistant to a given antibody if at least one of the entries of its corresponding state vector is non-zero.

We define an individual-based stochastic birth-death model \cite{wilkinsonStochasticModellingSystems2019a} to capture the competitive dynamics of different HIV variants within a population. This dynamic model will allow us to predict the distribution of rebound times under any combination of antibodies. 

We assume that a viral strain of type $a$ can undergo one of three processes: birth, death and mutation to another type $b$ with rates $\beta_a$, $\delta_a$, and $\mu_{a\to b}$, respectively:
\begin{align*}
\text{birth}:\quad [a] 
&\overset{\beta_a}{\longrightarrow} 2[a] \\
\text{death}:\quad [a] 
&\overset{\delta_a}{\longrightarrow}  * \\
\text{mutation}:\quad [a] 
&\overset{\mu_{a\to b }}{\longrightarrow}  [b]\\
\end{align*}

We specify  an intrinsic fitness $f_a$ for a given variant $a$, 
	defined as the growth rate of the virus in the absence of neutralizing antibody or competition. Since bNAbs target highly vulnerable regions of the virus,	we expect that   HIV escape mutations to be intrinsically deleterious for the virus and to confer a fitness cost relative to the susceptible viral variants  prior to the infusion of bNAbs. Assuming that fitness cost of escape is additive across sites and background-independent, we can express the fitness of a variant as,  
	$f_a  =  f_0 - \sum_{i} \Delta_{i} \rho_{i}^{a} $, where $\Delta_i$ is the cost associated with the presence of an escape mutation at site $i$ of variant $a$ (i.e., for $\rho_i^a =1$).

We assume that growth is self-limiting via a competition for host T-cells. This competition enforces a carrying capacity, which sets the steady-state population size $N_k$. Competition is mediated through a competitive pressure term $\phi =  \frac{\sum_a N_a f_a}{N_K}$ which attenuates the net growth rate $\gamma_a$ so that $\gamma_a = f_a - \phi$. At the carrying capacity, the competitive pressure equals the mean population fitness $\phi=\overline{f}$,
	making the net growth rate of the population zero.

The net growth rate of a variant $a$ is given  by its birth rate minus the death rate: $\gamma_a = \beta_a - \delta_a$.
We  assume that the total rate of events (i.e., the sum of birth and death events) is equal for all types, i.e.,  
	$\lambda = \beta_a + \delta_a, \,\forall a$.
Assuming that $\lambda$ is constant is to be agnostic about the mechanism of a fitness decrease, 
	attributing fitness loss equally to (i) an increase in  the death rate, and (ii) a decrease in the birth rate.

Because the absolute magnitude of $\beta$ and $\delta$ asymptotically converge in the continuum limit for a surviving population, i.e., $\lim_{N \rightarrow \infty}  \beta / \delta = 1$, it is impossible to distinguish between (i) and (ii) in the continuous limit.
Choosing constant $\lambda$ simplifies both theoretical calculations and the simulation algorithm.

This leads to the following equations for the birth and  the death rates:
\begin{align}
 \beta_i &= \frac{\lambda + (f_i - \phi)}{2} &  \delta_i &= \frac{\lambda - (f_i - \phi)}{2} \label{eq.rates}
 \end{align}
In the presence of an antibody, birth is effectively halted for susceptible variants, resulting in birth and death rate for a susceptible variant $s$,
\begin{align}
 \beta_s &= 0 &  \delta_s &= r
 \end{align}
so that  the susceptible phenotype decays at rate $r$.

We assume that mutations occur independently at each site,
\begin{align}
\mu_{a\to b} = 
	\begin{cases}
		\mu_{i}   &\text{if  } \rho^{a}-\rho^{b} = 1_{i}\\
		\mu^{\dagger}_{i}    &\text{if  } \rho^{a}-\rho^{a} = -1_{s}\\
		0     &\text{otherwise}\\
	\end{cases}
\end{align}
where $1_s$ is the vector which has only one non-zero entry at site $i$, and  $\mu_i$  and $\mu_i^\dagger$ are the forward the backward mutation rates at site $i$, respectively.

We characterize the state of a population by vector $\bm{n} = (n_1, \hdots n_M)$, where $n_a$ is the number of type $a$ variants within the population. The individual-based birth-death model introduced above specifies the stochastic dynamics of a population state over time. Using the concept of chemical reactions, suitable for Gillespie algorithm \cite{wilkinsonStochasticModellingSystems2019a, gillespieExactStochasticSimulation1977}, we can determine the propensity $a_r(\bm{n})$ for a given reaction $r$ (i.e., birth, death, or mutation) in a population of state $\bm n$, which in turn determines the rate at which the reactions occur (eq.~\ref{eq.rates}). We denote the resulting change in the state of a population  due to reaction $r$ by  $\bm{\nu}_r$. Taken together, the impact of the reactions in the birth-death model can be summarized as,
\begin{align}
 \begin{tabular}{l | | l l l l}
    \hline
        Reaction & & Rate parameter& Propensity  & State change  \\
        			&&				&  $a_r(\bm{n})$ & $\bm{\nu}_r$\\ \hline \hline
    Birth & & $\beta_a = \frac{\lambda + (f_i - \phi)}{2}$  &  $n_a\beta_a$& $ +\hat{e}_i$  \\ 
   Death & & $\delta_a = \frac{\lambda - (f_i - \phi)}{2}$&  $n_a \delta_a$ & $ -\hat{e}_i$ \\ 
 Mutation & & $\mu = \mu_{a\to b} $  &  $n_a \mu_{a\to b}$ & $+\hat{e}_b - \hat{e}_a$ \\ 
   && $(\mu^\dagger = \mu_{b \to a})$ & 			& 
            \end{tabular}
\label{eq:reactions}
            \end{align}
where $\hat{e}_i$ is a vector of size $M$ equal to size of  the population state vector, in which the $i^{th}$ element equal to one and the rest are zero. For example, 
a mutation reaction $\mu(a\to b)$ destroys a variant $a$ and creates a variant $b$, resulting in the following change in the state vector,
\begin{align}
    \bm{\nu}_{\mu(i\rightarrow j)} = - \hat{e}_i + \hat{e}_j 
\end{align}

The reactions in eq.~\ref{eq:reactions} specify a Master equation for the change in the probability of the population state $p(\bm n)$,
\begin{align}
    \dot{p}(\bm{n}) = \sum_r a_r(\bm{n} - \bm{\nu}_r) p(\bm{n}- \bm{\nu}_r) - a_r(\bm{n}) p(\bm{n})
\end{align}
where $\bm{n} = \sum_i n_i \hat{e}_i$ is the state vector. Using a Kramers-Moyal expansion~\cite{riskenFokkerPlanckEquationMethods1996}, we arrive at a Fokker-Planck approximation for the change in the probability distribution of the population state $p(\bm n)$,

\begin{align}
    \frac{\d}{\d t}{p}(\bm{n}) &= 
    \left[ \sum_r \left( \frac{1}{2}\sum_{i,j} \frac{\partial}{\partial n_i} \frac{\partial}{\partial n_j}   \nu^{i}_r \nu^{j}_r a_r(\bm{n}) -\sum_i\frac{\partial}{\partial n_i}   \nu^{i}_r a_{r}(\bm{n})  \right) \right]  p(\bm{n})
\end{align}

We can  identify the drift (i.e., the deterministic force) and diffusion tensors of the Fokker-Planck operator:
\begin{align}
 \bm{b}(\bm{n}) &= \sum_r \bm{\nu}_r a_r(\bm{n}) &  \bm{\Sigma}(\bm{n}) &= \sum_r \bm{\nu}_r^2 a_r (\bm{n})
\end{align}

To better demonstrate the structure of this birth-death operator,  consider a bi-allelic case (e.g. susceptible and resistant) with a 2-dimensional state vector, $\bf n =  \begin{pmatrix} n_0 \\
n_1 \end{pmatrix}$. The drift and the diffusion tensors associated with this process follow, 
\begin{align}
\bm{b}(\bm{n}) &= \begin{pmatrix} f_0 - \frac{n_0 f_0 + n_1 f_1}{N_k} - \mu & \mu^\dagger \\ \mu &  f_1 - \frac{n_0 f_0 + n_1 f_1}{N_k} - \mu^\dagger \end{pmatrix} \cdot  \bm{n} \\
\label{eq:sigma}\bm{\Sigma}(\bm{n}) &= \lambda \begin{pmatrix} n_0 & 0\\ 0&n_1\end{pmatrix} +\mathcal O(\mu)
\end{align}
Note that in eq.~\ref{eq:sigma} we neglect the stochasticity  due to mutations since the magnitude of the associated noise is much smaller than the noise due to the birth and death (i.e., genetic drift). 

Of special interest for our analysis is the steady-state density of frequencies, which we use to describe the initial state of the population (before treatment) and to infer selection intensity. In the steady state, the population is fluctuating around carrying capacity $\sum_a n_a \approx N_k$ and we can represent the population state via allele frequencies $x_a = n_a/N_k$. In the simple  case of a bi-allelic  problem, the equilibrium allele frequency distribution $P_\text{eq}(x)$ follows the Wright-equilibrium distribution \cite{crowIntroductionPopulationGenetics2010} with modified rates,
\begin{align}
P_\text{eq}(x) = \frac{1}{Z} \frac{ e^{\frac{2N_k }{\lambda} (f_1 - f_0)  x}\,\, (1-x)^{\frac{2N_k }{\lambda}\mu^\dagger} x^{{\frac{2N_k }{\lambda}}\mu}}{(1-x)x} \equiv \frac{1}{Z(\sigma,\theta,\theta^\dagger)} \frac{ e^{ -\sigma x} (1-x)^{\theta^\dagger} x^{\theta} }{(1-x)x} 
\label{P_eq:SI}
\end{align}  
where $Z$ is the normalization factor,  $f_1$ is the intrinsic fitness of the variant of interest, $f_0$ is the fitness of the competing variant, $\mu$ and $\mu^\dagger$ are the forward and backward mutation rates, $N_k$ is the carrying capacity, and $\lambda$ is the total rate of events in the birth-death process, which sets a characteristic time scale over which the impact of selection and mutations can be measured. In this case, we can define an ``effective population size" that sets the effective size of a bottleneck and the natural time scale of  evolution as $N_e= N_k/\lambda$, and specify a scaled selection factor $\sigma = N_e s = N_e ( f_0- f_1)$, and scaled forward  mutation and backward mutation rates (diversity) $\theta = 2N_e \mu$, and $\theta^\dagger = 2N_e \mu^\dagger$. The normalization factor is given by, 
\EQ
Z\equiv Z(\sigma,\theta,\theta^\dagger) = {\cal B} (\theta,\theta^\dagger)\,   _{1} F_1(\theta,\theta+\theta^\dagger, -\sigma)
\label{eq.normEq:SI}
\EE
where $_{1} F_1(\cdot)$ denotes a Kummer confluent hypergeometric function and ${\cal B} (\theta,\theta^\dagger) = \frac{\Gamma[\theta] \Gamma[\theta^\dagger]}{\Gamma[\theta+\theta^\dagger]}$ is the Euler beta function.

\subsection{Extinction Probability}
The logistic dynamics describing a patient's viremia over time  in eq.~\ref{eq:logisticpiecewiseSI} is the deterministic approximation to the underlying birth-death process. However, the resistant population can also go extinct due stochastic effects, which  in turn contribute to  the probability of late rebound in a population. To capture this effect, we derive an approximate closed form expression for the probability of extinction.

Using the standard birth-death process generating function theory~\cite{allenIntroductionStochasticProcesses2010} the probability $P(\text{extinct}|n_i) $ that a population consisting of  $n_i$ resistant variants of type $i$ go extinct can be expressed as, 
\begin{align}
P(\text{extinct}|n_i) = \left( \frac{\delta_i}{\beta_i} \right)^{n_i}.\label{eq.extProb_1_SI}
\end{align}
To characterize the probability of extinction for a population of  size $N_k$ with pre-treatment fraction of $i^{th}$ resistant variants $x_i$, we can convolve the extinction probability in eq.~\ref{eq.extProb_1_SI} with a Binomial probability density for sampling $n_i$ resistant variants from  $N_k$ trials. Given that the pre-treatment fraction of resistant variants  is small $x_i\ll 1$ and $N_k$ is large, this Binomial distribution can be well approximated by a Poisson distribution, $\text{Poiss}(n_i; N_kx_i)$ with rate $N_k x_i$, resulting in an extinction probability,
\begin{align}
\nonumber P(\text{extinct}|x_i) &= \sum_{n_i} \text{Poiss}(n_i; N_kx_i)\left( \frac{\delta_i}{\beta_i} \right)^{n_i} \\
\nonumber	&=  \exp(-N_k x_i) \sum_{n_i} \frac{(N_k x_i)^{n_i}}{n_i !}  \left( \frac{\delta_i}{\beta_i} \right)^{n_i} \\
	&= \exp \left( -N_k \frac{\beta_i - \delta_i}{\beta_i} x_i \right)
\end{align}
Using the expressions for the growth in the absence of competition, $\beta_i - \delta_i = \gamma_i = f_i$ (since $\phi = 0$), and assuming that fitness  is small relative to the total rate of birth and death events $f_i\ll \lambda$, we can use the approximation $\beta_i  = (\lambda + f_i) /2 \approx \lambda/2$, to arrive at,
\begin{align}
P(\text{extinct}|x_i) 
	& \approx \exp\left( \frac{-2 N_k f_i}{\lambda} x_i  \right) = \exp \left( - \frac{x_i}{x_\ext} \right) \label{eq:extinctprob}
\end{align}
Where the characteristic escape threshold $x_\ext$ can be written in terms of concrete genetic observables,
\begin{align}
x_\ext \equiv  \frac{\lambda}{ 2 N_k f_i}  = \frac{\mu_{\ts}}{f_i} \theta_{\ts}^{-1}.
\label{eq.xesc_SI}
\end{align}
Fig.~3 shows that this threshold can well separate the fate of stochastic  evolutionary trajectories, simulated with relevant parameters for intra-patient HIV evolution.

\subsection{Numerical simulations of the birth-death process}
To treat the full viral dynamics including mutations, and transient competition effects, we can exactly simulate the viral dynamics defined by our individual based model. Below are the key steps in this simulation.\\

\noindent{\em Population initialization.} 
At the starting point, we set the population size  (i.e., the carrying capacity in the simulations) $N_k$ as a free parameter chosen to be large enough to make discretization effects small.
The population is then evolved through time using an exact stochastic sampling procedure (the Gillespie algorithm \cite{gillespieExactStochasticSimulation1977}).
 Simulating the outcome of this stochastic evolution generates the distribution of rebound times and the probability of late rebound---the key quantities related to treatment efficacy.

The input to our procedure is a list of antibodies for which we specify (i)  the escape mediating sites for each antibody, and the (invariant) quantities describing (ii)  the site-specific  cost of escape $\frac{\sigma}{\theta_{\ts}}$, and  (iii) the forward and backward mutation rates $(\frac{\theta}{\theta_{\ts}}, \frac{\theta^\dagger}{\theta_{\ts}})$. To simulate the trial outcome for each patient, we use the neutral population diversity  $\theta_{\ts}$ directly inferred from the patients (see Sec.~\ref{sec:diversity}). From this, we construct the list of $L$ site parameters (concatenated across all antibodies) for selection and diversity: $\sigma_{1:L}$, $\theta_{1:L}$, $\theta^{\dagger}_{1:L}$.

We assume that at the start of the simulation, populations are in the steady state and that the potential escape sites are at linkage equilibrium. The approximate  linkage equilibrium  assumption is  justified  since the distance between  these escape sites along the HIV genome is greater than the characteristic recombination length scale $\approx 100\text{bp}$ of the virus \cite{Zanini:2015gg}. As  a result, we draw an independent frequency $x_i$ from the  stationary distribution $P_\eq(x|\sigma_i,\theta_i,\theta^\dagger_i)$ in eq.~\ref{P_eq:SI} to describe the state of a give site $i$, and use these frequencies to construct the initial viral genotypes $\rho^{v}$ for each virus $v$ in our initial population (Algorithm~\ref{alg:popinit}). 
 In simulations, we show that this assumption does not bias our results even when $\theta_{\ts}$ is fluctuating and recombination is absent (Section~\ref{sec:robustness} and Fig.~\ref{Fig:S6}).

\begin{algorithm}[h!]
\caption{Population initialization} \label{alg:popinit}
\begin{algorithmic}
\Procedure{PopulationInitialization}{$N_k, \sigma_{1:L}, \theta_{1:L}, \theta^{\dagger}_{1:L}$}\\ 
\CommentG{$N_k$, the carrying capacity, determines the initial population size at equilibrium. $\theta$, and $\sigma$ are the parameters defining the equilibrium values of the population state. Returns the initial vector of genotypes}
\For{$i \in 1:L$}
	\State $x_i \sim P_\eq(x|\sigma_i,\theta_i,\theta^\dagger_i)$ 
\EndFor
\For{$v \in 1:N_k$}
	$\rho^{v}_i \sim \text{Bernoulli}(x_i)$ 
\EndFor \\
\Return $\rho^{1:N_k}$ 
\EndProcedure
\end{algorithmic}
\end{algorithm}

To sample from the stationary  distribution itself, we define a novel Gibbs-sampling procedure \cite{gemanStochasticRelaxationGibbs1984a} for generating the allele frequencies of the escape variants for the initial state of the population $x \sim P_\eq(x|\sigma,\theta,\theta^\dagger)$ (eq.~\ref{P_eq:SI}). To characterize this procedure, we expand the exponential selection factor  $e^{\sigma (1-x)}$ in the original distribution, which results in,
 \begin{align}
\nonumber P_\eq(x|\sigma,\theta,\theta^\dagger)  &= \frac{e^{-\sigma}}{Z(\sigma,\theta,\theta^\dagger)} e^{\sigma (1-x)}  \frac{ x^{\theta}  (1-x)^{\theta^\dagger} }{x(1-x)}\\
\nonumber &=  \sum_{k=0}^\infty \frac{ e^{-\sigma} }{Z(\sigma,\theta,\theta^\dagger)}    \frac{\sigma^k}{k!}  x^{\theta} \frac{(1-x)^{\theta^\dagger+k} }{x(1-x)}\\
& \equiv  \sum_{k=0}^\infty Q_\eq(x,k |\sigma,\theta,\theta^\dagger)
\end{align}
Here, $Q_\eq(x,k |\sigma,\theta,\theta^\dagger)$ is a joint distribution over $(x,k)$, and the desired distribution over the allele frequency $x$ can be achieved by marginalizing the joint distribution over the discrete variable $k$. We can also express the conditional distributions for $x$ and $k$ as, 

\begin{align}
Q_\eq(x | k, \sigma,\theta,\theta^\dagger) &= \frac{Q_\eq(x,k |\sigma,\theta,\theta^\dagger)}{\int \d x \, Q_\eq(x,k |\sigma,\theta,\theta^\dagger)} =  \text{Beta}(x; \theta , \theta^\dagger+k)\\
Q_\eq(k | x, \sigma,\theta,\theta^\dagger) &= \frac{Q_\eq(x,k |\sigma,\theta,\theta^\dagger)}{\sum_k Q_\eq(x,k |\sigma,\theta,\theta^\dagger)} =  \text{Poisson}(k ;(1-x) \sigma)
 \label{eq.GibbsSample_SI}
\end{align}
We use these conditional distributions to define a joint Gibbs sampler for $Q_\eq$.
We summarize the resulting $(x,k) \sim Q_\eq(x,k |\sigma,\theta,\theta^\dagger)$ in the joint Gibbs sampler in Algorithm~\ref{alg:sample}. This chain mixes extremely quickly and avoids calculation of the hypergeometric function for the normalization factor (eq.~\ref{eq.GibbsSample_SI}), which is computationally costly (Algorithm~\ref{alg:sample}).

\begin{algorithm}[h!]
\caption{Gibbs Sampler for Allele Frequencies}\label{alg:sample}
\begin{algorithmic}
\Procedure{EqulibriumSampler}{$\sigma, \theta, \theta^{\dagger} | \text{Samples} , \text{BurnIn}$} 
\CommentG{Generates a stream of non-independent but rapidly mixing samples $X \sim p(x|\sigma, \theta, \theta^{\dagger})$. Default $\text{BurnIn}$ is $10$.}
\State $N \gets \text{Samples} + \text{BurnIn}$
\State $K_0 \gets \text{Round}(\sigma)$
\For{$n \in 1:N$}
\State $X_{n+1} \sim \text{Beta}(\theta, \theta^\dagger+K_{n})$ \CommentG{Sample the mutant fraction}
\State $K_{n+1} \sim \text{Poisson}(\sigma (1-X_{n+1}))$ \CommentG{Sample the auxiliary parameter} 
\EndFor \\
\Return $X_{\text{BurnIn} : N}$ \CommentG{Return only the mutant frequency $X$ part of the chain (marginalize over $k$)}
\EndProcedure
\end{algorithmic}
\end{algorithm}

\noindent {\em Simulation of the evolutionary process.} 
We use a Gillespie algorithm to simulate the evolutionary process, where we break up the reaction calculation into two parts: randomly choosing a viral strain $\rho_i$ from the population and then determining whether it reproduces or dies based on its fitness $f_i$ and escape status (Algorithm~\ref{alg:cap}). 

\begin{algorithm}[h!]
\caption{Population time step}\label{alg:cap}
\begin{algorithmic}
\Procedure{EvolvePopulation}{$t, \rho^{1:N} \vert \lambda, \gamma, r,$} \\
\CommentG{Acts on a time $t$ and a list of $N$ genotypes $\rho^{1:N}$. Inherits dependency on other parameters from the fitness function $F(G)$ and the $\text{Mutate}(G)$ operator which depend on $\Delta_{1:L}, \mu_{1:L}, \mu^{\dagger}_{1:L}$ and $\gamma$ and the population diversity measure $\theta_{\ts}$.} 
\State $\phi \gets \frac{1}{N} \sum_i F(g_i)$
\State $t' \gets  t + \frac{\text{RandExp()}}{ \lambda N}$ \CommentG{Advance time}
\State $i \sim  \text{Rand}(1:N)$
\State $G \gets \rho^{i}$
\If{$\text{IsEscaped}(G)$} \CommentG{If the virus is escaped} 
	\State $D \sim \text{Bernoulli}(\frac{\lambda- F(G) + \phi}{2\lambda})$ \CommentG{Determine if the virus dies ($D = \text{true}$)  or lives ($D = \text{false}$).} 
	\If{$D$}
		\State $N' \gets N-1$
		\State $\rho^{1:N'} \gets \rho^{1:\tilde{i}:N}$ \CommentG{Delete genotype at position $i$} 
	\Else
		\State $N' \gets N+1$
		\State $\rho^{1:N'} \gets \text{Append}(\rho^{1:N}, G) $ \CommentG{Duplicate genotype at position $i$}
	\EndIf
\Else \CommentG{\,If the virus is neutralized}
	\State $D \sim \text{Bernoulli}(\frac{r}{\lambda})$ \CommentG{ remove it at the appropriate rate}
	\If{$D$}
		\State $N' \gets N-1$
		\State $\rho^{1:N'} \gets \rho^{1:\tilde{i}:N}$ \CommentG{Delete genotype at position $i$} 
	\EndIf
\EndIf
\State $j \sim  \text{Rand}(1:N')$ \CommentG{Choose a random virus to mutate}
\State $\rho^j \gets \text{Mutate}(\rho^j)$ \CommentG{Apply mutation operator with intensity $\mu/\lambda$} \\
\Return $(t', \rho^{1:N'})$ \CommentG{Return the new time and the new population.}
\EndProcedure
\end{algorithmic}
\end{algorithm}

\subsection{Determining the simulation parameters of the birth-death process from genetic data}
We set the intrinsic growth rate (fitness) of the wild-type virus, in the absence of competition to be $\gamma = (3 \text{ days})^{-1}$, consistent with intra-patient doubling time of the virus ~\cite{garciaVirologicalImmunologicalConsequences2001, ioannidisDynamicsHIV1Viral2000, garciaDynamicsViralLoad1999}.  We infer the neutralization rate $r$ by fitting the viremia curves Fig.~\ref{Fig:S1} in the trials under study, and use the averaged decay rate $r=0.31$ for simulations, fitted using eq.~\ref{tab:decay_rates}. For the absolute mutation rate $\mu_{\ts}$ (per nucleotide per day) we use $1.1 \times 10^{-5}$ which is the average of the reported values for transitions per site per day from ref.~\cite{Zanini:2017in}. 
Using the  covariance of neutral diversity in two-fold and four-fold synonymous sites, we determine the  transition/transversion diversity ratio to be $\theta_{\ts}/\theta_{\tv} = 7.8$ (Fig.~\ref{Fig:S4} and Section~\ref{sec:diversity}). We use these values to  determine the forward and backward mutation rates $\mu_s$ and $\mu_s^\dagger$ for each site (Section \ref{sec:mut_target}).

Generally the trial patients show a larger viral diversity at the start of the trial compared to the patients enrolled in the high-throughput study of ref.~\cite{Zanini:2015gg} (Fig.~2).  We account for differences in the genetic makeup of the patients enrolled in the trial by directly estimating the viral diversity $\theta_{\ts}$ from the neutral site-frequency data of patients, before the start of the trial. The estimated viral diversity $\theta_{\ts}$, coupled with the mutation rate $\mu = 1.1 \times 10^{-5} \text{ /day / nt}$, and the total rate of birth and death events in the viral population   $\lambda$, set the carrying capacity $N_k$ for a given individual,
\begin{align}	
	N_k = \frac{\lambda  }{2  \mu_{\ts}}  \theta_{\ts}.
	\label{eq:noise_popsize}
\end{align}

It should be noted that the value of $N_k$, as the number of viruses in our individual-based simulations,  
	is not related to the maximal viral load in the viremia measurements (i.e., steady state copy number per ml) 
	as this relationship depends on the microscopic details of the population dynamics.

The total rate of birth and death events $\lambda=\beta+\delta$ tunes the amount of  stochasticity, i.e.,  more events cause noisier dynamics. Notably, stochasticity can be linked to the size of the population  $N_k$, which is directly coupled to $\lambda$ (eq.~\ref{eq:noise_popsize}). We set the value of $\lambda$ self-consistently by requiring that  the minimum frequency of a variant in our simulations $x_\text{min} = 1/N_k  = \frac{2 \mu_{\ts}}{\lambda} \theta_{\ts}^{-1}$ to be smaller than the  escape threshold $x_\text{esc} = \frac{\mu_{\ts}}{\gamma} \theta_{\ts}^{-1}$ due to stochasticity (eq.~\ref{eq.xesc_SI}). We set $\lambda = 2 \text{day}^{-1}$ so that $x_\text{min} = \frac{1}{3} x_\text{esc}$. Increasing $\lambda$ results in an in crease in the  size of population $N_k$ in our simulations, which is computationally costly, without qualitatively changing the statistics of the rebound trajectories.

\section{Inference of diversity, mutational target size, and selection from genetic data}
\subsection{Inference of mutation rates and the neutral diversity within a population.}
\label{sec:diversity}
Previous work has indicated an order of magnitude difference between  the rate of transitions (mutations within a nucleotide class) and transversions (out-class mutations) in HIV~\cite{nielsenStatisticalMethodsMolecular2006,Zanini:2017in, theysWithinpatientMutationFrequencies2018b, federSpatiotemporalAssessmentSimian2017}.  Therefore, to infer the neutral diversity parameter $\theta_{\ts}$, we also account for the differences between transition and transversion rates.

Consider the set of sequences sampled from a patient's viral population at a particular time. Two neutral alleles that are linked by a symmetric mutational process $\mu_{1\rightarrow2} = \mu_{2\rightarrow1}$  have a simple count likelihood.
The probability to see allele 1 with multiplicity   $n$  and allele 2 with multiplicity $m$  is given by a binomial distribution $\Binom(n,m|x)$ with parameter $x$ denoting the probability for occurrence of allele  1,  convolved with the neutral biallelic frequency distribution $P_\eq(x|\sigma=0,\theta)$ from eq.~\ref{P_eq:SI}. Using this probability distribution,  we can evaluate the log-likelihood $\L(\theta|n,m)$ for the neutral diversity $\theta$ given the observations $(n,m)$ for the multiplicities of the two alleles in the population,
\begin{align}
\label{eq:equilibrium_likelihood}
\L(\theta|n,m) &= \log \int\, \d x \, \Binom(n,m|x) \times P_\eq(x| \sigma=0,\theta) \\
&= \log \int \d x \, {n+m\choose n} x^n (1-x)^m\,\times \frac{x^{\theta-1}(1-x)^{\theta-1}}{Z(\theta)} 
\end{align}

\begin{figure*} [t!]
    \centering
	\includegraphics[width=0.9\textwidth]{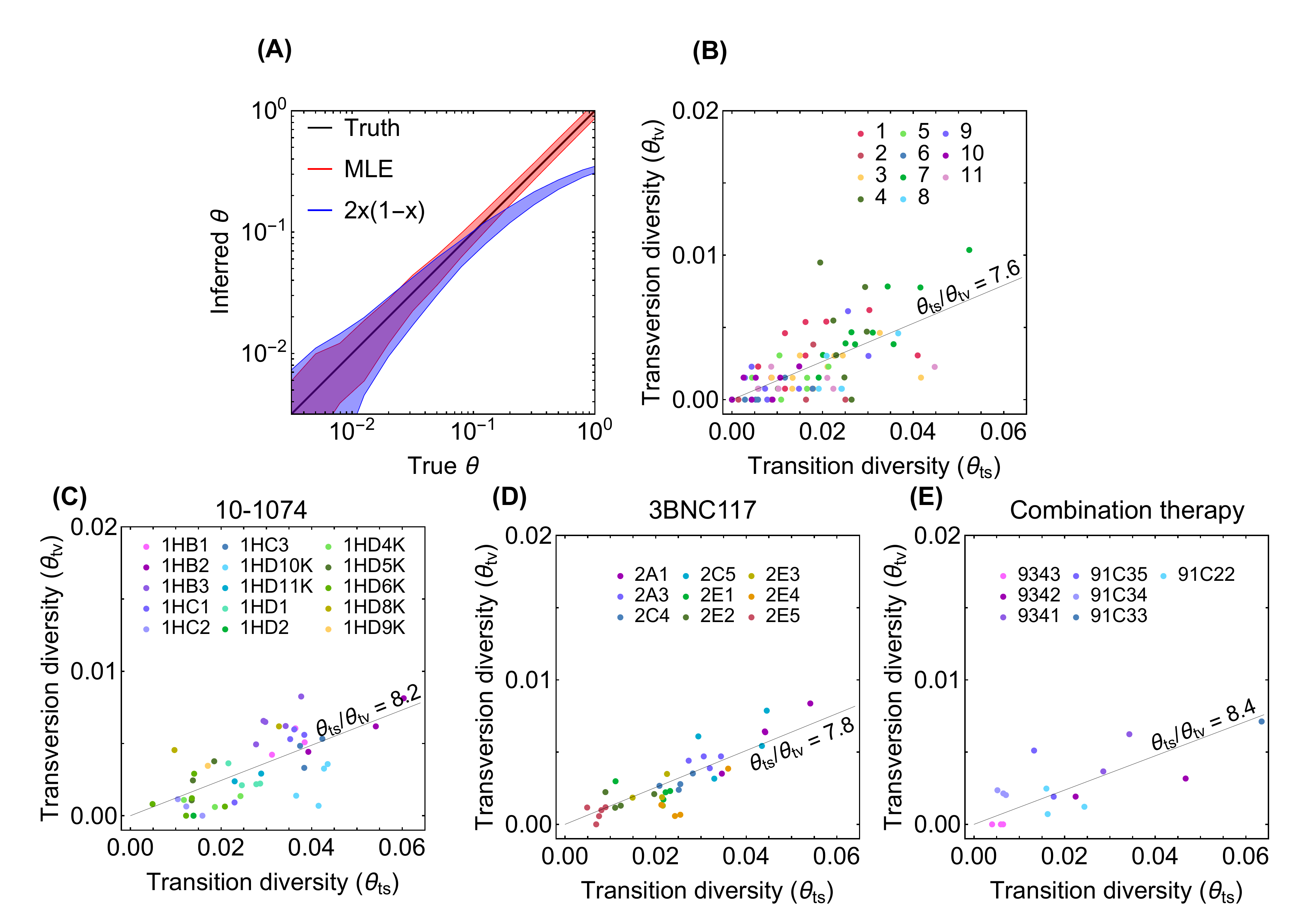}
    \caption{
    {\bf Inference of diversity from genetic data.}  
    {\bf (A)} The inter-decile ranges of diversity estimates for $10^3$ simulations of genomes of $10^2$ independent neutral sites are shown. The true values of the diversity used for simulations are shown in black.  Inter-decile maximum likelihood estimate (MLE) for the diversity $\theta$ from eq.~\ref{eq:equilibrium_likelihood} are shown in red, and the observed variance of allele frequency $x$ as a  measure of diversity $\theta =2 x (1-x)$ is shown in blue.    
    The inferred diversity associated with transversions $\theta_\tv$ is shown against that of the transition $\theta_\ts$ for patients (colors) enrolled in  {\bf (B)} the longitudinal study with high throughput HIV sequence data~\cite{Zanini:2015gg}, and in the bNAB trials with {\bf (C)} 10-1074~\cite{Caskey:2017el}, {\bf (D)} 3BNC117~\cite{Caskey:2015hm}, and {\bf (E)} combination of 10-1074 and 3BNC117~\cite{bar-onSafetyAntiviralActivity2018}. We find that all four diversity distributions share similar ratios $\theta_\ts/\theta_\tv$, indicated in each panel. \\
  }
   \label{Fig:S4}
\end{figure*}

To estimate the transition diversity, we only use  two-fold synonymous sites, and treat each site  independently but with a shared diversity parameter $\theta_{\ts}$.
For example, consider neutral variations for two amino acids glutamine and phenylalanine.  The third position in a codon for  both of these amino acids are two-fold synonymous, as the two possible codons for glutamine are CAG and CAA, and for phenylalanine are TTT, and TTC. Now consider that in the data a conserved glutamine has $n = 3$  G's and $m = 97$  A's in the third codon position, and a conserved phenylalanine has  $n=10$ T's and $m=90$ C''s at its  third codon position. In this case, the combined log-likelihood for the shared diversity parameter is $\L(\theta|\text{data}) = \L(\theta|3,97) + \L(\theta|10,90)$. Extending to all of sites in the {\it env} protein, the maximum-likelihood estimator for the transition diversity $\theta_{\ts}$ can be evaluated by maximizing the likelihood summed over all conserved two-fold synonymous sites, 
{\small
\begin{align}
\theta^*_{\ts} = \arg \max_{\theta_{\ts}} \sum_\text{two-fold sites} \L(\theta_{\ts}|n = n_A +n_T ; m =  n_G + n_C)
\end{align}} 
In Fig.~\ref{Fig:S4} (panel A) we show that the maximum likelihood estimation method described above has better properties than the more commonly used estimator of the variance $\overline{x(1-x)}$  \cite{stoddartGenotypicDiversityEstimation1988}.

In a similar way, the likelihood for the transversion $\theta_{\tv}$ is determined from polymorphic data at all conserved four-fold synonymous sites. One such example is the third position in a glycine codon, where (GGT, GGC, GGA, GGG) translate to the same amino acid. The maximum-likelihood estimator for the transversions is
{\small
\begin{align}
\theta^*_{\tv} = \arg \max_{\theta_{\tv}} \sum_\text{four-fold sites} \L(2\theta_{\tv}|n = n_G +n_T ; m =  n_C + n_A)
\end{align}}
The factor of $2$ in the argument of the likelihood accounts for the multiplicity of mutational pathways, e.g. from a $G$ nucleotide there are two transversion possibilities, $G\rightarrow C$ and $G\rightarrow A$ for  moving from one allele to the other \cite{kimuraEstimationEvolutionaryDistances1981}.

Using this likelihood approach, we can infer the neutral diversities $\theta^*_{\ts}$ and $\theta^*_{\tv}$ for each patient at each time point from the polymorphism in two-fold and four-fold synonymous sites. To characterize the ratio of transition to transversion rates,  we use linear regression on the entire patient population and sample history and infer a constant ratio $\mu_{\ts}/\mu_{\tv} = \theta_{\ts}/\theta_{\tv} = 7.8$ (Fig.~\ref{Fig:S4}B). In Fig.~\ref{Fig:S4}, we also show that the estimate for this ratio  is relatively consistent across different data sources, produced  even by different sequencing technologies. The previously reported relative rate of transitions to transversions,  based on the estimates of sequence divergence along phylogenetic trees of HIV-1 is $\mu_{\ts}/\mu_{\tv}= 5.6$~\cite{Zanini:2017in}, which is similar to our maximum likelihood estimate.

\subsection{Inference of mutational target size for each bNAb}
\label{sec:mut_target}
The nucleotide triplets which encode for amino acids at an escape site 
	undergoe substitutions which can change the amino acid type
	and create an escape variant.
The changes in the state of an amino acid  codon can be modeled as a Markov jump process 
	and can be visualized as a weighted graph where 
	the nodes represent codon states, and edges represent single nucleotide substitutions
	linking two codon states (Fig.~1D).
In our mutational model, these edges have weights associated with either the mutation rates for transitions $\mu_{\ts}$ or transversions $\mu_{\tv}$. We call this the {\em codon substitution graph}.

The codon states can be clustered into three distinct classes: (i) codons which are fatal $F$,  (ii) wild-type (i.e., susceptible to neutralization by the bNAb) $W$, and escape mutants ( i.e. resistant to the bNAb) $M$.
We expect the escape mutants to be at a selective disadvantage compared to  the resistant wild-type, and that the most common escape codons to be those which are adjacent to wild-type states.

The mutational target size is determined by the density of paths from the wild-type $W$ to the escape mutants $M$. 
\begin{align}
\mu &= \frac{1}{|W|} \sum_{c\in W ; d \in M}  [c-d = \text{ts}] \mu_{\ts} + [c-d = \text{tv}]  \mu_{\tv} \\
\mu^\dagger &= \frac{1}{|M|} \sum_{c \in W ; d \in M}[c-d = \text{ts}] \mu_{\ts}+  [c-d = \text{tv}] \mu_{\tv} 
\end{align}
The functions $[c-d = \text{ts}]$ or $[c-d = \text{tv}]$ are 1 when the two codons are separated by a transition or transversion, and are zero otherwise.  Note that since we only have an estimate for the ratio of the transition to transversion rates $\mu_{\tv}/\mu_{\ts}$,
	 we can only determine the scaled mutational target sizes, $\hat{\mu}  = \mu/\mu_{\ts}$ and $\hat{\mu}^\dagger =  \mu^\dagger/\mu_{\ts}$, which are sufficient for inference of selection in the next section. The full list of mutational target sizes inferred for the  bNAbs in this study are  shown in Fig.~\ref{tab:full_antibody_data}.

When discussing the mutational target size of escape from a given bNAb, we refer to the total mutation rate from the susceptible (wild type) to the escape variant as,
\begin{align}
\mu = \sum_{i\in \text{ esc. sites} }\mu_{s\to i}
\end{align}
where the sum runs over all the mediating escape sites $i$, and  can be interpreted as the average number of accessible escape variants. In the strong selection regime, we can write the frequency of escape mutants as,
\begin{align}
x_\text{mut} &\approx \sum_i x_i = \sum_i \frac{\mu_{s\to i}}{\Delta_i} \approx  \frac{\mu}{\Delta_\text{hm}} 
\end{align}

\subsection{Inference of selection for escape mutations against each bNAb}

Here, we  develop an approximate likelihood approach to infer  the selection ratio $\hat{\sigma} = \frac{\sigma}{\theta_{\ts}}$, using the high-throughput sequence data from bNAb-naive HIV patients from ref.~\cite{Zanini:2017in}. The quantity $\hat{\sigma}$ is a dimensionless ratio which is independent of the coalescence timescale $N_e$, and therefore, represents a stable target for inference.

We assume that the probability to sample $m$ escape mutants and $s$ susceptible (wild-type) alleles at a given site  in the genome of  HIV in a   population sampled from a patient at a given time point follows a binomial distribution $\text{Binom}(m,s |x)$,  governed by the underlying  frequency  $x$ of the mutant allele. In addition, the frequency $x$ of the allele of interest  itself is drawn from the equilibrium distribution $P_\eq(x|\sigma,\theta,\theta^\dagger)$ (eq.~\ref{P_eq:SI}), governed by the diversity  $\theta_{\ts}$ inferred from the neutral sites, the estimated mutational target sizes $\hat{\mu} = \mu/\mu_{\ts}, \hat{\mu}^\dagger =  \mu^\dagger/\mu_{\ts}$, and the unknown selection ratio $\hat{\sigma} = \sigma/\theta_{\ts}$. As a result, we can characterize the probability $ P(m,s | \theta_{\ts} , \hat{\mu}, \hat{\mu}^\dagger, \hat{\sigma})$ to sample $m$ escape mutants and $s$ susceptible-type alleles, given the scaled selection and diversity parameters as,   
\begin{align}
\nonumber
P(m,s | \theta_\ts, \hat{\sigma} ) & = P(m,s | \sigma = \hat{\sigma} \theta_{\ts} , \theta = \hat{\mu} \theta_{\ts}, \theta^\dagger = \hat{\mu}^\dagger \theta_{\ts}) \\
\nonumber&= \int \d x\, \text{Binom}(m,s |x) P_\eq(x| \sigma, \theta, \theta^\dagger )\\
&=  \frac{1}{Z(\sigma,\theta,\theta^\dagger)} \binom{s+m}{s} \int \, \d x  \frac{e^{-\sigma x} x^{\theta + m} (1-x)^{\theta^\dagger+s}}{x(1-x)} \label{eq:Pmutwt_SI}
\end{align}
Here, $Z(\sigma,\theta,\theta^\dagger)$ is a confluent hypergeometric function of the model parameters that sets the normalization factor for the allele frequency distribution $P_\eq(x)$ (eq.~\ref{eq.normEq:SI}).
It should be noted that  the viral population is in fact out of equilibrium, due to constant changes in immune pressure evolution of the B-cell and T-cell populations. Although we are ignoring these significant complications, we later use the same equilibrium distribution in a consistent way to generate standing variation in simulations. For the model to make accurate predictions, it is not necessary that the equilibrium model be exactly correct,
	but only that it is rich enough to provide a consistent description for the distribution of mutant frequencies observed across viral populations.

We will use the probability density in eq.~\ref{eq:Pmutwt_SI} to define a  log-likelihood function in order to infer the scaled selection $\hat{\sigma}=\sigma/\theta_\ts $ from data.  To do so, we first express the logarithm of this probability density  as, 
\begin{align}
\nonumber \log P(m,s | \theta_{\ts}, \hat{\sigma}) & = \log P(m,s | \sigma = \hat{\sigma} \theta_{\ts} , \theta = \hat{\mu} \theta_{\ts}, \theta^\dagger = \hat{\mu}^\dagger \theta_{\ts}) \\
\nonumber &= \log Z(\sigma,\theta+ m,\theta^\dagger +s) - \log Z(\sigma,\theta,\theta^\dagger) + \text{const.}\\
\label{eq:betaAve_SI}&=\log \E\left[ e^{- \sigma x}\right]_{ \text{Beta}(\theta +  m, \theta^\dagger + s)}-  \log \E\left[e^{- \sigma x} \right]_{\text{Beta}(\theta, \theta^\dagger)} +\text{const.} 
\end{align}
where the constant factors ($\text{const.}$) are independent of selection,  and $\E[\cdot]_{\text{Beta}(\cdot)}$ denotes the expectation  of the argument over a $\text{Beta}$ distribution with parameters specified in the subscript. The expression in eq.~\ref{eq:betaAve_SI} implies that we can evaluate the likelihood of selection strength by computing the difference  between the logarithms of the expectation for $e^{-\sigma x}$ over allele frequencies drawn from two neutral distributions (Beta distributions), with parameters $(\theta,\theta^\dagger)$ and  $(\theta+m,\theta^\dagger+s)$, respectively. This approach is  more attractive as it would not require direct evaluation of the confluent hypergeometric functions for the normalization factors in eq.~\ref{eq:Pmutwt_SI}. Estimating these normalization factors is computationally intensive for large values of  $\sigma$, since many terms in the underlying hypergeometric series should be taken into account to stably compute them. However, evaluating the expectations via sampling  from these two neutral distributions  has the disadvantage that it is subject to  variations across simulations. We reduce the variance of our estimate of $\log P(m,s|\sigma,\theta,\theta^\dagger) $ in eq.~\ref{eq:betaAve_SI} by using a mixture-importance sampling scheme \cite{owenSafeEffectiveImportance2000} with the details shown in Algorithm~\ref{alg:loglike}.

\begin{algorithm}[h!]
\caption{Importance sampled log-likelihood given a single datapoint}\label{alg:loglike}
\begin{algorithmic}
\Procedure{SigmaLikelihood}{$\hat{\mu},\hat{\mu}^\dagger, m, s, \theta_\ts, N$} 
\CommentG{Takes mutant $m$ and susceptible $s$ counts for a particular site at a single timepoint. Returns an approximate log-likelihood function, $l(\hat{\sigma}) = \log P(m, s |\theta_{\ts},\hat{\sigma}) +c$ up to an additive constant. $\textsc{SigmaLikelihood}$ is a {\em closure} that returns a one-parameter function. We used $N = 10^3$ samples.} 
\State $\theta \gets \hat{\mu} \theta_\ts$
\State $\theta^\dagger \gets \hat{\mu}^\dagger \theta_\ts$
\For{$i \in 1:N$ }
\State $x_i \sim \text{Beta}(\theta,\theta^\dagger)$ \CommentG{Sample from the neutral distribution}\\
\State $w_i \gets \frac{{\cal B}(\theta + m, \theta^\dagger+s)}{{\cal B}(\theta,\theta^\dagger)} \frac{1}{x_i^m (1-x_i)^w } $  \CommentG{Importance weight ratio}
\State $y_i \sim  \text{Beta}(\theta + m, \theta^\dagger+s)$ \CommentG{sample from the neutral distribution, conditioned on the observations} 
\State $v_i \gets \frac{{\cal B}(\theta + m, \theta^\dagger+s)}{{\cal B}(\theta,\theta^\dagger)} \frac{1}{y_i^m (1-y_i)^w } $  \CommentG{Importance weight ratio}
\EndFor
\State $Z_0(\hat{\sigma}) := \frac{1}{N} \sum_i e^{\hat{\sigma} \theta_\ts x_i} \frac{1}{1+w_i} + \frac{1}{N} \sum_i e^{-\hat{\sigma} \theta_\ts y_i} \frac{1}{1+v_i} $ \CommentG{importance sampling mean}
\State $Z_1(\hat{\sigma}) := \frac{1}{N} \sum_i e^{\hat{\sigma} \theta_\ts x_i} \frac{1}{1+w_i^{-1}} + \frac{1}{N} \sum_i e^{-\hat{\sigma} \theta_\ts y_i} \frac{1}{1+v_i^{-1}}$  \CommentG{Note the inversion in the weighting factor compared to $Z_0$.}
\\
\Return $l (\hat{\sigma}) := \log Z_1(\hat{\sigma}) - \log Z_0(\hat{\sigma})$ \CommentG{Return a log-likelihood function. The same random variable realizations $x_{1:N}$ and $y_{1:N}$ are cached in memory and used for each function evaluation, making $l (\hat{\sigma})$ continuous and differentiable.}
\EndProcedure
\end{algorithmic}
\end{algorithm}

We use data collected across all time points and from all patients to infer reliable estimates for selection strengths. However, allele frequencies are correlated across time points within patients (Fig.~2B), and thus, sequential measurements are not independent data points.
Our estimates indicate a coalescence time of about $N_e\sim10^3$ days based on the estimates for the mutation rate  $\mu_\ts = 10^{-5}$ /nt/day, and the neutral diversity $\theta_\ts =  2 N_e  \mu_\ts=0.01 $. This coalescence time is much longer that the typical separation between sampled time points within a patient ($\sim 10^2$ days), suggesting that   sequential samples collected from  each individual in this data are correlated.  
Therefore, we treat each patient as effectively a single observation, using the time-averaged likelihood for the (scaled) selection factor $\hat \sigma$:
 \begin{align}
 \label{eq:timeaveraged}
{\cal L }(\hat{\sigma}) &= 
	\sum_{p} \frac{1}{T_p} \sum_t \log P(m_t, s_t |\theta_{\ts}(t), \hat{\sigma})
\end{align}
where $p$ and $t$ denote patient identity and sampled time points, respectively, and $T_p$ is the total number of time points sampled in patient $p$. We use  the likelihood in eq.~\ref{eq:timeaveraged} to generate samples from the posterior distribution for selection strengths under a flat prior, with a standard Metropolis-Hastings algorithm \cite{hastingsMonteCarloSampling1970a}.  Since the prior is constant, this procedure amounts to simply accepting or rejecting samples based on the likelihood ratio of eq.~\ref{eq:timeaveraged}.
We used the centered-normal distribution with standard deviation of $50$ ($\times \mu_\ts$ in absolute units) as the proposal density for the jumps in the Markov chain.

\section{Predicting trial outcomes from genetically informed evolutionary models}

\paragraph{Predicting rebound times}
 We expect different distributions of patient outcomes depending on whether they have been recently infected and thus have relatively low viral diversity, or whether their infection is longstanding with a diverse viral population. To construct the distribution of initial population diversities $\theta_{\ts}$ for simulating trial outcomes, we apply the $\theta_{\ts}$ inference procedure (eq.~\ref{eq:equilibrium_likelihood}) to pre-treatment sequence datasets available from the clinical trials under consideration.
In Fig.~2D this set of pre-trial $\theta_\ts$ is compared to the longitudinal in-patient diversity.
We used random draws from the inferred $\theta_\ts$ values for patients to generate $\theta_\ts$ for simulations.

We found that there was considerably more viral escape and non-responders in our simulations than in the observed data as shown in Fig.~\ref{Fig:S7}A.
This is {\em in addition} to the fact that the patients were screened to have only susceptible variants to the antibodies used in trials \cite{Caskey:2015hm,Caskey:2017el,bar-onSafetyAntiviralActivity2018}. In theory, there should be zero non-responders, as such patients should have been excluded by screening. The over-prediction of {\em both} non-responders and late rebounds is a signature of undercounting the effective diversity of the viral populations.

The failure of both screening and our naive prediction in undercounting the diversity in the viral population can be explained by an effective viral reservoir.
Viral variants which mediate rebound can come from compartments such as including bone marrow, lymph nodes, and organ tissues, and can be genetically distinct from those sample from the plasma T-cells during screening \cite{chaillonHIVPersistsThroughout2020a,wongTissueReservoirsHIV2016,chunHIVinfectedIndividualsReceiving2005}.
This reservoir of viral diversity can reappear in plasma after infusion of a bNAb and could  in part contribute to  treatment failure \cite{avettand-fenoelFailureBoneMarrow2007, shanReactivationLatentHIV12013, sharkeyEpisomalViralCDNAs2011, tagarroEarlyHighlySuppressive2018}. \\

\paragraph{Determining patient diversity enhancement due to latent reservoirs}
\label{sec:reservoir_factor} 
We model the effect of  reservoirs as a simple inflation of the diversity observed by a multiplicative factor $\xi$.
We fit $\xi$ directly to trial observations, using a disparity-based  approach by minimizing an empirical divergence estimator \cite{ekstromAlternativesMaximumLikelihood2008} between the observed and simulated data. To do so, we characterize the Hellinger distance \cite{lindsayEfficiencyRobustnessCase1994, simpsonMinimumHellingerDistance1987} between the true distribution of rebound times $P(t)$ and the rebound times  $Q(t|\xi)$ generated by simulations with a given reservoir factor $\xi$,
\begin{align}
D_H(P(t)||Q(t|\xi) ) = \int \d t\, (Q(t|\xi)^\frac{1}{2} - P(t)^\frac{1}{2} )^2 \approx \sum_{(i)}^{n_q} (Q_{(i)}(\xi)^\frac{1}{2} - P_{(i)}^\frac{1}{2})^2
\end{align}
Algorithm~\ref{alg:disparity} defines the procedure that we use to estimate  the Hellinger distance $D_H(Q(t)||P(t|\xi) )$. Specifically, we use $n_q$ quantiles of the observed data $x_{(i)} \sim Q$ to partition the space of observations into discrete outcomes 
\begin{align}
Q_{(i)}(\xi) &= \int \d t \, P(t|\xi) 
\left[ t_{(i)} \leq t <  t_{(i+1)}\right] &
P_{(i)} &=  \frac{1}{n_q}.
\end{align}
where $P_{(i)}$ is a constant by construction, and $P_{(i)}(\xi)$ is estimated by simulations, and $\left[ \cdot \right] $ is the Iverson bracket \cite{grahamConcreteMathematicsFoundation1989} (Algorithm~\ref{alg:disparity})
\begin{align}
\left[ B \right] &= \begin{cases}
1 \text{ if } B = \text{true}\\
0 \text{ if } B = \text{false}
\label{eq:Iverson}
\end{cases}
\end{align}

To simulate data for this analysis, we generate $S$ rebound times ($T_{1:S}$)  by simulations, given the scaled diversity values $\xi \theta_\ts$. 
We then find the optimal value $\xi^*$ by minimizing the disparity with the observed rebound times $t_{1:p}$ by brute-force search,
\begin{align}
\label{eq:disparityR}
	R(\xi|t_{1:p}) &= \sum_\text{trials} \textsc{ReboundDisparity}(t_{1:p},T_{1:S}| \xi)\\
	\xi^* &= \arg \min_\xi R(\xi|t_{1:p}) \label{eq:disparity_min}
\end{align}
Here, $\textsc{ReboundDisparity}$ is the function defined by Algorithm~\ref{alg:disparity}; see ref.~\cite{ekstromAlternativesMaximumLikelihood2008} for details.
We find the optimal reservoir factor to be $\xi^* = 2.1$, which we use in subsequent therapy prediction. The disparity over various values of $\xi$ for different trials is shown in Fig.~\ref{Fig:S7}. \\

\begin{algorithm}[h!]
\caption{Rebound-time Disparity \label{alg:disparity}}
\begin{algorithmic}
\Procedure{ReboundDisparity}{$t_{1:P}, T_{1:S}$} \\
\CommentG{Takes the observed rebound times $t_{1:P}$ ($\sim Q$) from a set of trial patients and simulated late rebound times  $T_{1:S}$ $\sim P$) and returns a disparity estimator.} \\
\CommentG{First estimate the probabilities in the truncated observation categories}
\State $P_{(NR)} \gets \frac{1}{P}\sum_p [ t_p < 1 \text{ day} ]$ \CommentG{Count the fraction of non-responders in trial}
\State $Q_{(NR)} \gets \frac{1}{S} \sum_s [ T_s < 1 \text{ day} ]$ \CommentG{ ... and in simulation}
\State $P_{(LR)} \gets \frac{1}{P} \sum_p [ t_p \geq 56 \text{ days} ]$ \CommentG{Count the fraction of late rebounds in trial}
\State $Q_{(LR)} \gets \frac{1}{S} \sum_s [ T_s \geq 56 \text{ days} ]$ \CommentG{... and in simulation}\\
\CommentG{Then construct a histogram over the continuous data-points (i.e. $t \in (0,56]$) and estimate the probability in each bin}
\State $t_{1:P'} = \text{SortAscending}(Filter[1 \leq t <  56](t_{1:p}))$ \CommentG{Select only the observed rebound times}
\State $t_0 \gets - \infty$ and $t_{P'+1} \gets \infty$
\For{$p \in 1:P'$}
\State $Q_{(p)} \gets \frac{1}{P}$ \CommentG{By design, each histogram bin contains one observed data point, and gets $1/P$ mass}
\State $P_{(p)} \gets \frac{1}{S} \sum_s [ \max(1, \frac{1}{2} (t_{p-1} + t_p))  \leq T_s < \min(56, \frac{1}{2} (t_{p+1} + t_p))  \text{ days} ]$ \CommentG{Use the midpoints of adjacent points to construct the boundaries of histogram bins, and determine probability mass in each bin.}
\EndFor \\
\Return{$P \sum_{i \in NR,LR,1:P'} (Q_{(i)}^{1/2} - P_{(i)}^{1/2})^2$} \CommentG{Return the discretized estimate of the Hellinger distance, scaled by the number of patients}
\EndProcedure
\end{algorithmic}
\end{algorithm}

\paragraph{Simulating outcomes of the clinal trails}
Given the reservoir-corrected estimate of the diversity $\xi \theta$ and the posterior samples for selection factors $\hat{\sigma}$, we now summarize how we simulated the outcome of clinical trials.

For a given bNAb, we draw the selection factor at each of the escape mediating sites from the corresponding Bayesian posterior on $\hat{\sigma}$; the posterior distributions are shown in Fig.~4A, and summarized in Table~\ref{tab:full_antibody_data}. We also use the mutational target size  (forward $\mu$ and backward $\mu^\dagger$ rates) associated with each of the escape mediating sites of a given bNAb; see Table~\ref{tab:full_antibody_data}. The result can be summarized in a  mutation / selection matrix  $\hat{M}_a$ for a given bNAb $a$, where each column corresponds to an escape mediating site $i$ against the bNAb,
\begin{align}
\hat{M}_a &= \begin{pmatrix}
\hat{\mu}_1 &\hdots& \hat{\mu}_i\\
\hat{\mu}_1^\dagger & \hdots & \hat{\mu}_i^\dagger\\
\hat{\sigma}_1 &\hdots  & \hat{\sigma}_i\\
\end{pmatrix}
\end{align}
The elements of the matrix $\hat{M}_a$ are the scaled mutation and selection factors, i.e., $\hat{\mu}_i = \mu_i/\mu_\ts,\, \hat{\mu}^\dagger_i = \mu_i/\mu_\ts,\, \text{and } \hat\sigma_i = \sigma_i/\theta_\ts$, where the absolute value of mutation rate is set to  $\mu_{\ts}= 1.11 \times 10^{-5}\, /\text{nt}/\text{day}$ from ref.~\cite{Zanini:2017in}.

For each patient in our simulated trial, we then draw diversity  $\theta_{\ts}$ from the patient pool, and scale it by our fitted $\xi = 2.1$, resulting in patient-specific selection and mutation factors,
\begin{align}
\xi \times \theta_{\ts} \times  \hat{M}_a &= 
\begin{pmatrix}
{\theta}_1 &\hdots& {\theta}_i\\
{\theta}_1^\dagger & \hdots & {\theta}_i^\dagger\\
{\sigma}_1 &\hdots  & {\sigma}_i\\
\end{pmatrix} &
	\mu_{\ts} \times  \hat{M}_a &= 
\begin{pmatrix}
{\mu}_1 &\hdots& {\mu}_i\\
{\mu}_1^\dagger & \hdots & {\mu}_i^\dagger\\
{\Delta}_1 &\hdots  & {\Delta}_i\\
\end{pmatrix} 
\label{eq.SI_patient_mut_sel}
\end{align}
These parameters are then used to initialize the state of an HIV population within a patient according to Algorithm~\ref{alg:popinit}, and to determine the absolute rates in Algorithm~\ref{alg:cap} for the population evolution according to eq.~\ref{eq.SI_patient_mut_sel}. The decay rate is set to the fitted trial average of $r=0.31 \, \text{days}^{-1}$ (eq.~\ref{tab:decay_rates}).  The carrying capacity $N_k$ is set according to eq.~\ref{eq:noise_popsize}. This determines all parameters of the  birth-death process simulating the intra-patient evolution of HIV, which are used in Algorithm~\ref{alg:cap}.

We evolve a population through time until 56 days have elapsed since treatment, or until the escape fraction relative to the carrying capacity $x_t$ is above $0.8$ (Algorithm~\ref{alg:cap}). After $x_t >.8$ the evolution is governed by the deterministic equations, and the stochastic simulation ends. The rebound time $T$, defined as the intersection of the exponential envelope and the carrying capacity, can then be calculated analytically as,
\begin{align}
T = \frac{1}{\gamma} \log(1 + \exp(\gamma t) \frac{1-x_t}{x_t}).
\end{align}
The resulting distribution for rebound times are shown as model predictions in Fig.~3B-D.

Rebound times generated in this fashion were also used to estimate the probability of late rebound  to characterize the efficacy of a given bNAb in curbing viral rebound. The probability of late rebound was estimated from $10^4$ simulated patients. The interdecile quantiles (0.1 - 0.9) of early rebound probability over $200$ values of scaled selection coefficients $\hat{\sigma}$ drawn from the posteriors in Fig.~4A are shown in Fig.~4C.

\section{Model robustness}
\label{sec:robustness}

\paragraph{Effect of genomic linkage on the inference of selection}
In our inference of selection (eq.~\ref{eq:timeaveraged}, Fig.~4A), we assume that the escape-mediating sites are at linkage equilibrium and that the distribution of allele frequencies can be approximated by a  skewed Beta distribution (eq. \ref{P_eq:SI}), reflecting the equilibrium of allele frequencies. In reality, despite recombination, the HIV genome exhibits linkage effects, especially at nearby sites \cite{Zanini:2015gg}, and the viral populations experience changing selective pressures by the immune system~\cite{federClarifyingRoleTime2021,theysWithinpatientMutationFrequencies2018b,Nourmohammad:2019ij}, and the transient population bottlenecks during therapy~\cite{Feder:2016bc}.

To test the limits on the validity of our inference procedure, we applied it to {\em in-silico} populations generated by full-genome forward-time simulations (Algorithm~\ref{alg:cap}) in the presence and absence of recombination. To do so, we considered an ensemble of ten patients with 100 genomes sampled at ten time points, and used two diversity parameters $\theta_{\ts} = 0.01$ and $\theta_{\ts} = 0.1$, to cover the range reflected in patient data (Fig.~2D,~\ref{Fig:S6}). 

One relevant scenario to consider is the impact of other selected sites in the genome on  the distribution of alleles at the escape mediating sites against bNAbs. The sites under a strong constant selection are likely to be already fixed (or at a high a frequency) at their favorable state in the population. However, the strong selection on a large fraction of antigenic sites in HIV can be thought as time-varying, due to the changing pressure imposed by the immune system or therapy. To capture this effect,   we simulated whole genome evolution in which  linked sites were under strong selection ($0.1 \times \text{growth rate}$), and where the sign of selection changed after exponentially distributed waiting times (i.e. as a Poisson process); this model of fluctuating selection has been used in the context of influenza evolution~\cite{Strelkowa:2012jo}, and for somatic evolution of B-cell repertoires in HIV patients~\cite{Nourmohammad:2019ij}.  
The resulted evolutionary dynamics in  this case can involve strong selective sweeps and clonal interference due to the continuous rise of beneficial mutations  (in the linked sites) within a population. 

To test the robustness of our selection inference, we evaluated the distribution of maximum likelihood estimates (MLEs) for the selection values  $\hat{\sigma} = \sigma / \theta_ts$ at the escape mediating sites, inferred from the ensemble of sequences obtain from  simulated data with linkage. Fig.~\ref{Fig:S6} shows that even for fully linked genomes (zero recombination) our MLE estimate of selection has little bias relative to the true values used in the simulations. Adding recombination into the simulations only further attenuates the effect of linkage (Fig.~\ref{Fig:S6}), making the estimates more accurate.

The reason that selective sweeps of linked beneficial mutations have only minor  effects on our inference of selection for the escape mediating sites  is two-fold: First, the primary mechanism by which selective sweeps change the strength of selection at linked sites is via a reduction in the effective population size \cite{hillEffectLinkageLimits1966, comeronHillRobertsonEffect2008}.
However, variations in the effective population size are already accounted for in our inference procedure: The selection likelihood in eq.~\ref{eq:timeaveraged} is conditioned on the measured neutral-site diversity, $\theta = 2 N_e \mu$. The change in the effective population size impacts the selection coefficient $\sigma = 2 N_e \Delta$ and the diversity $\theta=2 N_e \mu$  in the same way, and therefore, the  (scaled) selection parameter $\hat{\sigma} = \sigma/\theta$ that we infer from data remains insensitive to changes in the effective population size.

The second reason for the robustness of our selection inference to linkage  is due to the fact that a beneficial allele in a linked locus can appear on a genetic background with or without a susceptible variant, leading to the rise of either variants in the population.  As a results, the impact of such hitchhiking remains a secondary issue in inference of selection at the escape sites, for which an ensemble of populations from different patients with distinct  evolutionary histories of HIV is used.\\

\paragraph{Robustness  of selection inference to intra-patient temporal correlations among HIV strains}
To infer the selection effect of mutations from the longitudinal deep sequencing data of \cite{Zanini:2015gg}, we use time averaging of the likelihood (eq.~\ref{eq:timeaveraged}) to avoid conflating our results due to temporal correlations between the circulating alleles within patients (Fig.~2B). We can view this choice as being one choice among two extremes: (i) to treat each patient as effectively one independent data point so that all patients are given the same weight, or (ii) to treat each time point as independent, giving patients with more time points a higher weight.
These two choices correspond to different log-likelihood functions for the (scaled) selection factor $\hat \sigma$:

 \begin{align}
 \label{eq:tindependent}
{\cal L }(\hat{\sigma}) = \begin{cases}
	\sum_{p,t} \log P_p(m_t, s_t |\theta_{\ts}(t),\hat{\sigma}) & (\text{$t$-independent}) \\\\
	\sum_{p} \frac{1}{T_p} \sum_t \log P_p(m_t, s_t |\theta_{\ts}(t), \hat{\sigma}) & (\text{$t$-averaged})
\end{cases} 
\end{align}
where $T_p$ is the total number of time points from patient $p$, and $P_p(m_t, s_t |\theta_{\ts}(t), \hat{\sigma})$ is the probability to observe $m$ escape mutants, and $s$ susceptible variants at time $t$ in patient $p$, given the neutral diversity $\theta_{\ts}(t)$ and the scaled selection factor $\hat{\sigma}$. 

\begin{figure*} [t!]
    \centering
	\includegraphics[width=0.9\textwidth]{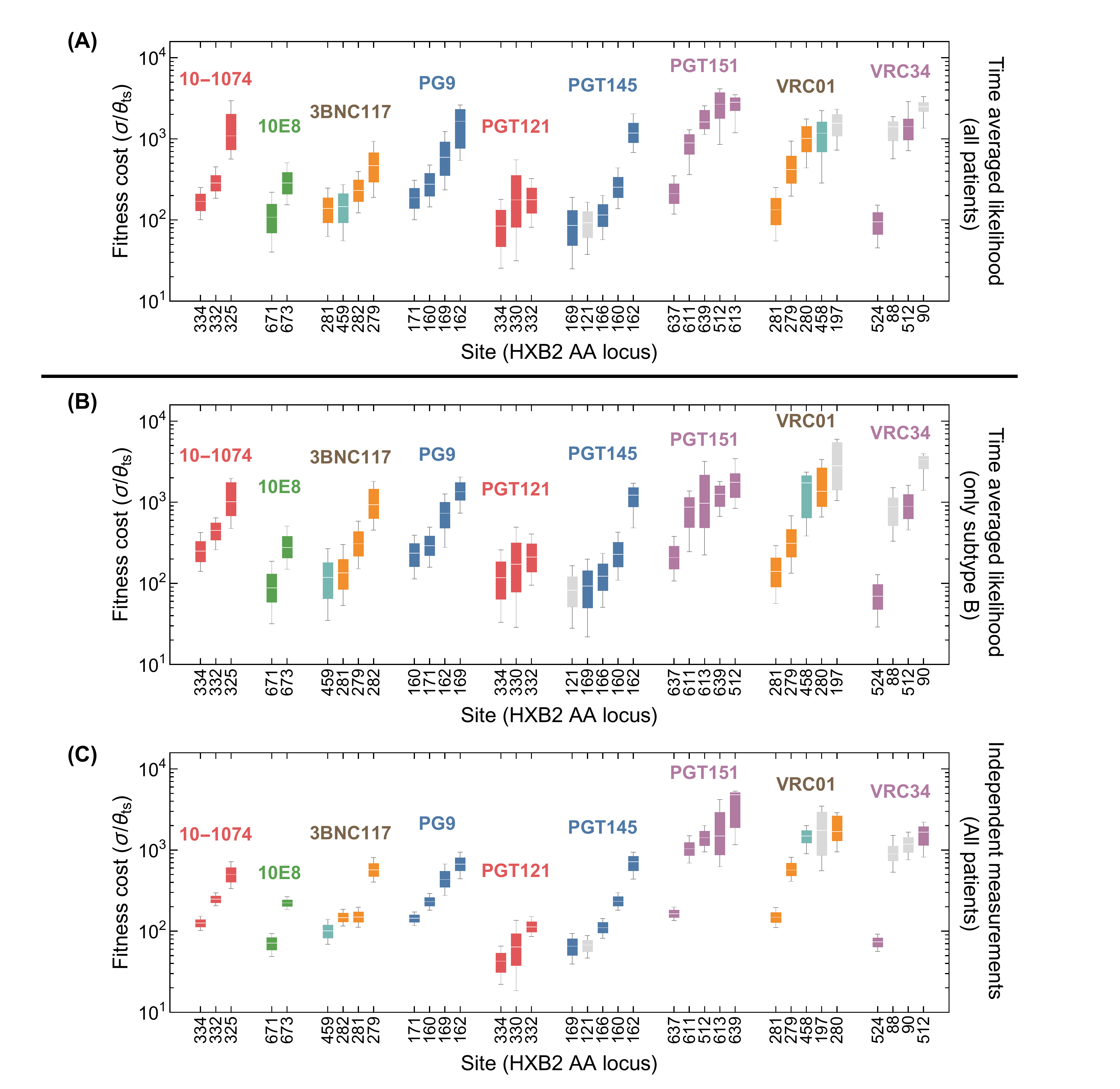}
    \caption{
    {\bf     Robustness of selection inference to correlations among HIV strains.} 
  Fitness cost for escape mediating sites of different bNAbs is shown based on the selection inference using {\bf (A)} the time-averaged likelihood (eq.~\ref{eq:tindependent} $t$-averaged) with data from all patients, as a reference (similar to Fig.~4A),  {\bf (B)} the time-averaged likelihood (eq.~\ref{eq:tindependent} $t$-averaged) with data from patients infected with HIV clade B only, and   {\bf (C)} independent time likelihood (eq.~\ref{eq:tindependent} $t$-independent) with data from all patients. Limiting the inference to patients infected with clade B virus in (B) result in minor changes in the inferred selection strength and a slight increase in uncertainties due to a smaller data. However, the general structure of the posteriors remain similar to (A). Treating all time points as independent in (C) increases the effective sample-size of the data and reduces model uncertainty. Nonetheless, except for  slight changes on the selection ordering of escape sites against 3BNC117, the median of the selection posteriors remain similar to (A).  The color code is similar to Fig.~4A and is determined by the {\em env} region associated with each site. 
}
   \label{Fig:S5}
\end{figure*}

\begin{figure*} [t!]
    \centering
	\includegraphics[width=1\textwidth]{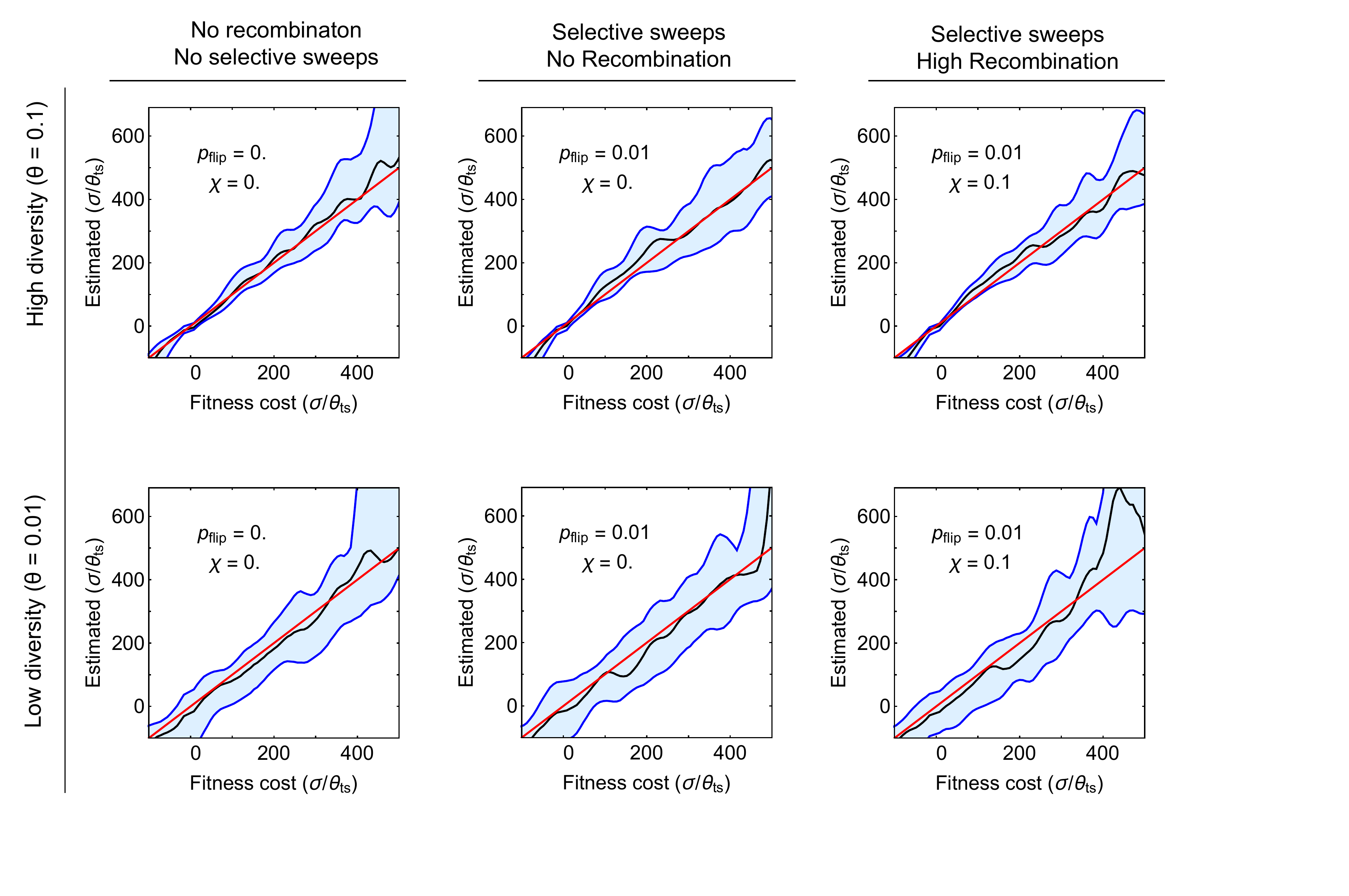}
    \caption{
    {\bf Robustness of selection inference to genetic linkage and hitchhiking.}
 Results of selection inference for simulated populations that carry escape mediating sites linked with other selected sites in the genome are shown.  
 The blue shading shows the interdecile ranges (0.1 -0.9 quantiles)  for the inferred maximum likelihood estimates (MLE) of  scaled selection $\hat{\sigma} = \sigma / \theta_\ts$ on the escape mediating sites versus the true values used in the simulations, for different genetic linkage effects (columns) and different values of population diversity (rows).  
Each inferred value is based on 10 realizations  of simulations ({\em in-silico} patients) for populations evolving over time, with samples taken at 10 time points spaced at $100$ days. Simulations are done on whole genomes of length $512$, in which $32$ sites under strong selection $0.02 f_0$ are equally spaced throughout the genome; $f_0$ is the base growth rate of $1.0 \text{ day}^{-1}$. Each day, the preferred allele in one of the $32$ sites is swapped (selection changes sign) with probability $p_\text{flip}$. The rate of selection fluctuation $p_\text{flip}$ and the recombination rate $\chi$ are given in each panel. The carrying capacity is set to  $N_k=5,000$, and the  event rate is $\lambda = 2 \text{ day}^{-1}$; see Algorithm~\ref{alg:cap}. To initialize, we let the populations equilibrate for $4 N_e$ generations, and then gather data for maximum likelihood inference of selection at the escape mediating sites. In all panels, the median MLE for selection  (black line) closely matches the true selection strength (redline). Quantile lines are smoothened by a Gaussian kernel. Variation is largest when diversity is low and when mutants are rare, which concords with the results from the Bayesian inference procedure.
} 
   \label{Fig:S6}
\end{figure*}

We find that both of these approaches result in similar posteriors for selection $\hat{\sigma}$ (Fig.~\ref{Fig:S5}A and C) although the $t$-averaged likelihood has a higher uncertainty due to fewer independent time points. Thus, our inference of selection is insensitive to the exact choice of the likelihood function given in eq.~\ref{eq:tindependent}, yet our time-averaged approach remains the more conservative choice between the two.\\

\begin{figure*} [t!]
    \centering
	\includegraphics[width=0.9\textwidth]{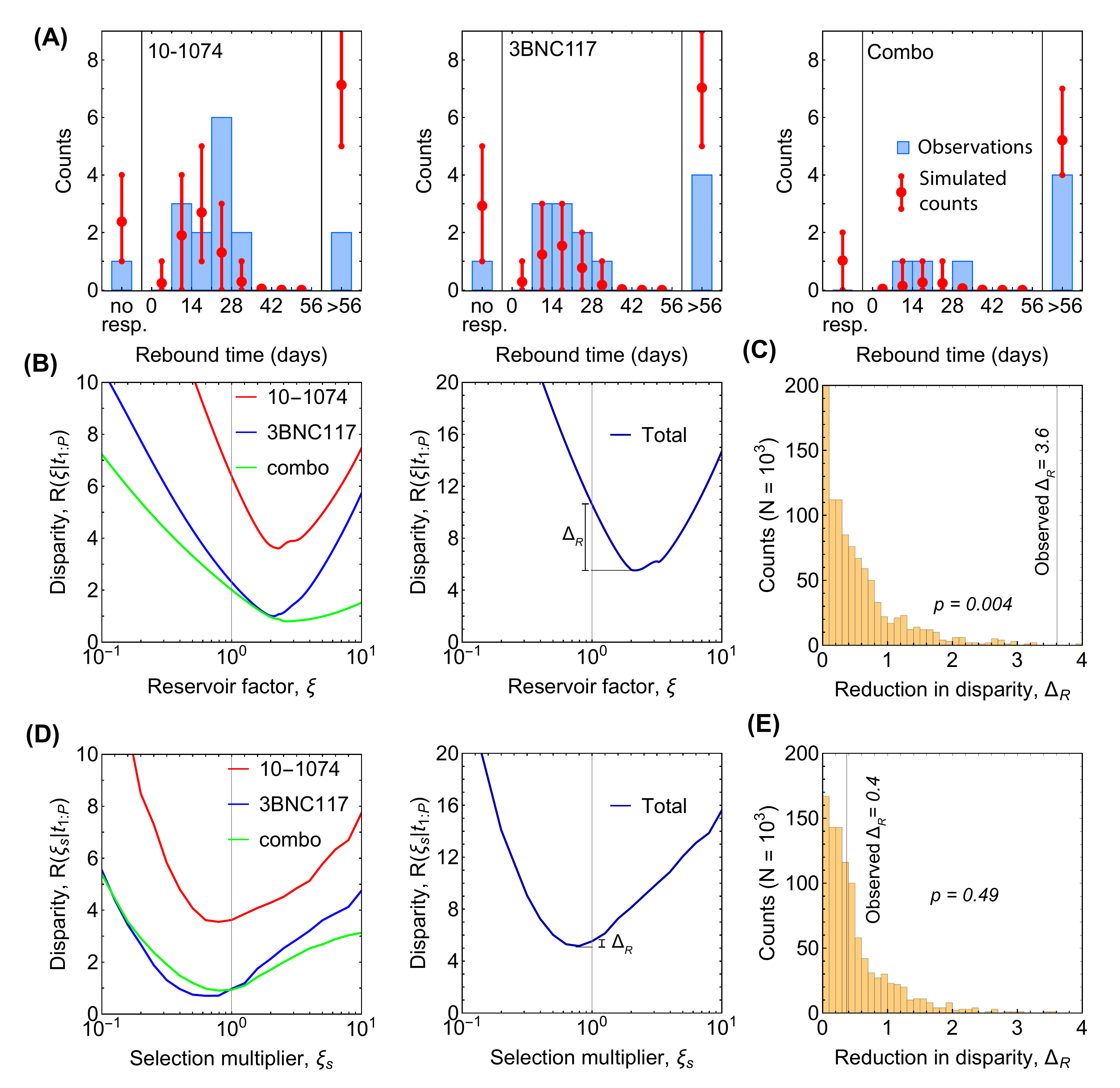}
    \caption{
    {\bf Minimum disparity estimation for adjustment of  diversity and selection.} 
    {\bf (A)} The predicted rebound times  without including  an increase in diversity due to reservoirs (i.e., $\xi_0=1$ in eq.~\ref{eq:SI_pvalue_disparity}) qualitatively matches  the data  from the three trials (histograms in each panel), with an overestimation in the number of non-responders and patients with late rebound. For comparison, the predictions with  reservoir-corrected diversity is  shown in Fig.~3A.
    {\bf (B)} The disparity $R(\xi|t_{1:p})$ (eq.~\ref{eq:disparityR}) is shown as a function of the reservoir factor $\xi$ for the three trials (colors; left panel). The total disparity over all trials is shown in the right panel, where $\Delta_R$ indicates the reduction in disparity by using the optimal reservoir factor $\xi^*=2.1$.     {\bf (C)} The distribution for the reduction in disparity $\Delta_R$ (eq.~\ref{eq:SI_DeltaR}) is shown for 1000 realizations of simulated data under the null hypothesis with $\xi_0=1$. The large reduction in disparity indicates that the null hypothesis  can be rejected with $\text{p-value} =0.004$ (eq.~\ref{eq:SI_pvalue_disparity}).  {\bf (D,~E)} Similar to (B,~C) but for assessing the sensitivity of the fit to rescaling of the selection strength  $\sigma/\theta_\ts$ by a multiplicative factor $\xi_s$. In this case the null hypothesis is that the inferred selection by the Bayesian procedure in eq.~\ref{eq:timeaveraged} requires no further rescaling for the model to fit the distribution of the rebound times (i.e., $\xi_s= 1)$. The large $\text{p-value} = 0.49$ shown in (E) indicates that the null hypothesis cannot be rejected and we have no statistical justification for adding an adjustment factor for selection inferred from untreated patients. 
}
   \label{Fig:S7}
\end{figure*}

\paragraph{Model of viral rebound with the reservoir-corrected  effects ($\xi = 1$)}
In eq.~\ref{eq:disparity_min} we introduced the reservoir factor $\xi^*=2.1$ to account  for the diversity of HIV that is not sampled from a patient's plasma prior to therapy, which resulted in a better fit of the rebound time distributions  (Fig.~3) compared to  a reservoir-free model  (Fig. \ref{Fig:S7}A). Here, we quantify the importance of the reservoir factor  with a statistical test on the null hypothesis, $\xi_0 = 1$. Specifically, we perform a hypothesis test to test the necessity of using an inflated diversity  $\xi\,\theta_\ts$ relative to using the bare diversity observed in pre-trial sequence data $\theta_\ts$. To do so, we construct a disparity-based test statistic \cite{ekstromAlternativesMaximumLikelihood2008}, which is analogous to the likelihood ratio test statistic. 

Recall that the optimal reservoir factor $\xi^*=\arg \min_\xi R(\xi|t_{1:p})$ was obtained  by minimizing the disparity function $ R(\xi|t_{1:p})$ across measurements of rebound times $t_{1:p}$ from all the $p$ patients in data (eq.~\ref{eq:disparity_min}). We can estimate the test statistic for the reduction in disparity between the null hypothesis, $\xi_0 = 1$ and the fitted reservoir factor  $\xi^*$ as,
\begin{align}
\Delta_R(t_{1:p})  = R(\xi_0 |t_{1:p}) - \min_\xi  R(\xi | t_{1:p}).
\label{eq:SI_DeltaR}
\end{align}
We can then determine the p-value by estimating the quantile of the observed test statistic $\Delta_R(t_{1:p})$ relative to that inferred from  the distribution of $\Delta_R(T_{1:p})$ obtained from simulations under null hypothesis $\xi_0 = 1$~\cite{fisherStatisticalMethodsScientific1956}. Specifically,
\begin{align}
\label{eq:SI_pvalue_disparity}
\text{p-value} = \expect{ T_{1:p} | \xi_0 } \Large( [ \Delta_R(T_{1:p}| \xi_0 ) > \Delta_R(t_{1:p}) ] \Large)
\end{align}
where  $\expect{ T_{1:p} | \xi_0}(\cdot) $ denotes the expectation over the rebound times $T_{1:p}$ obtained from  $1000$ realizations of simulated populations each with $p$ patients, and under the null hypothesis $\xi_0 = 1$. $\left[ \cdot \right] $ is the Iverson bracket  that takes value $1$ when its argument is true and $0$, otherwise (eq.~\ref{eq:Iverson}). The observed $\Delta_R$ (eq.~\ref{eq:SI_DeltaR}), the distribution of simulated values of $\Delta_R(T_{1:p}| \xi_0 )$ under the null hypothesis,  and the resulting $\text{p-value}= 0.004$ are shown in Fig.~\ref{Fig:S7}B,C.

It should be noted that here we use the disparity measure  because the corresponding likelihood function for the reservoir factor  is inaccessible through forward simulations of populations. However, a general analogy exists between our approach and the more commonly used likelihood approach. Specifically, in an analogous likelihood-ratio test, the test statistic $ \Delta_L = \max_\xi  \log p(\xi | t_{1:p}) - \log p(\xi_0 | t_{1:p})$ would be asymptotically $\chi^2$-distributed with one degree of freedom under the null hypothesis~\cite{fisherStatisticalMethodsResearch1954}, and the quantile under the null-hypothesis (p-value) would be estimated by inverting the $\chi^2$ cumulative distribution function (i.e. a $\chi^2$ test).  \\

\paragraph{Robustness of selection inference to  strains from different  clades of HIV}
The longitudinal deep sequencing data of \cite{Zanini:2015gg} is collected from 11 patients, 9 of whom are infected with clade B strains of HIV-1, which is the dominant clade circulating in  Europe \cite{spiraImpactCladeDiversity2003}. All of the clinical trials we considered~\cite{Caskey:2015hm,Caskey:2017el,bar-onSafetyAntiviralActivity2018} are from patients carrying clade B strains. For the results presented in the main text, {we included all the 11 patients in our analysis}. 
Here, we test wether our inference of selection is sensitive to the choice of including or excluding non-clade B patients in our analysis.
 We therefore repeated our inference procedure for selection by excluding the two non-clade B patients.
Fig.~\ref{Fig:S5}A,B shows a strong agreement between the Bayesian posterior  for selection factors in the two cases, with a slight increase in uncertainty for the case with only clade B patients. This increased uncertainty is related to the reduction in sample size by excluding the non-clade B patients from data. Nonetheless, the richness of the intra-patient diversity makes the inference robust to the exclusion of one or two patients.\\

\paragraph{Robustness of predictions for trial efficacy to the inferred values of selection strength}
How sensitive are the outcomes of our predictions for the rebound time distributions (Fig.~1) to the exact values of inferred selection strengths we used for our simulations?
We addressed this question by performing a disparity analysis similar to that for the diversity $\theta$ in eq.~\ref{eq:SI_DeltaR}. Specifically, we assessed whether we might  need to rescale our inferred selection strength $\sigma/\theta_\ts$ by a multiplicative factor $\xi_s$ (Fig.~\ref{Fig:S7}D,E). 

In contrast to diversity, the reduction in disparity  for adjustment of selection   with a factor $\xi_s$ is small (Fig.~\ref{Fig:S7}D) and not statistically significant ($\text{p-value}=0.49$; Fig.~\ref{Fig:S7}E), and could be attributed to count noise. Still,  we cannot discount the possibility that selection was slightly overestimated, possibly  due to the effect of compensatory mutations in linked genome, the interplay between the reservoir and the inference of selection, or other biological factors. Nonetheless, in absolute terms, the null hypothesis (i.e., $\xi_s=1$) cannot be rejected and we have no statistical justification for adding an adjustment factor for selection inferred from untreated patients.

\bibliographystyle{plos2}

\clearpage
\newpage
{\bf \large Supplementary Figures}\\

\begin{table*}[h!]
\scalebox{0.9}{
\begin{tabular}{lc|cc|cc|ccc|ccc}
     &      & & &     &               & \multicolumn{3}{c}{$ \sigma/\theta $ quantiles} & \multicolumn{3}{c}{$\sigma$ quantiles} \\ bNAb & site & susceptible AA & escape AA & $ \mu $ & $ \mu^\dagger $  & 0.1 & 0.5 & 0.9                         & 0.1 & 0.5 & 0.9 \\
\hline     
     \multirow{3}{*}{10-1074}
 & $325$ & DN & EGK & $0.76$ & $0.51$ & $5.6 \cdot 10^{2}$ & $1.1 \cdot 10^{3}$ & $2.9 \cdot 10^{3}$ & $8.8$ & $29$ & $93$\\
 & $332$ & N & DHIKSTY & $2.8$ & $0.4$ & $1.9 \cdot 10^{2}$ & $2.9 \cdot 10^{2}$ & $4.5 \cdot 10^{2}$ & $2.6$ & $7.1$ & $16$\\
 & $334$ & S & AFGINRY & $1.3$ & $0.44$ & $1 \cdot 10^{2}$ & $1.7 \cdot 10^{2}$ & $2.5 \cdot 10^{2}$ & $1.4$ & $4$ & $9.5$\\
\\
\multirow{2}{*}{10E8}
 & $671$ & KNS & RT & $0.41$ & $0.51$ & $38$ & $1.1 \cdot 10^{2}$ & $2.2 \cdot 10^{2}$ & $0.68$ & $2.5$ & $7.3$\\
 & $673$ & F & LV & $1.4$ & $0.47$ & $1.5 \cdot 10^{2}$ & $2.9 \cdot 10^{2}$ & $5.1 \cdot 10^{2}$ & $2.4$ & $7$ & $18$\\
\\
\multirow{4}{*}{3BNC117}
 & $279$ & DN & AHK & $0.33$ & $0.22$ & $1.9 \cdot 10^{2}$ & $4.7 \cdot 10^{2}$ & $9.3 \cdot 10^{2}$ & $3.4$ & $11$ & $31$\\
 & $281$ & AP & TV & $1.1$ & $1.1$ & $62$ & $1.4 \cdot 10^{2}$ & $2.5 \cdot 10^{2}$ & $0.97$ & $3.2$ & $8.6$\\
 & $282$ & KY & ENR & $1.2$ & $0.8$ & $1.2 \cdot 10^{2}$ & $2.3 \cdot 10^{2}$ & $4 \cdot 10^{2}$ & $1.9$ & $5.6$ & $14$\\
 & $459$ & G & DN & $0.5$ & $1$ & $54$ & $1.4 \cdot 10^{2}$ & $2.7 \cdot 10^{2}$ & $0.91$ & $3.3$ & $9.5$\\
\\
\multirow{4}{*}{PG9}
 & $160$ & N & KY & $0.4$ & $0.2$ & $1.5 \cdot 10^{2}$ & $2.8 \cdot 10^{2}$ & $4.8 \cdot 10^{2}$ & $2.3$ & $6.7$ & $17$\\
 & $162$ & ST & AIP & $1.2$ & $1.1$ & $5.4 \cdot 10^{2}$ & $1.6 \cdot 10^{3}$ & $2.6 \cdot 10^{3}$ & $9.5$ & $34$ & $98$\\
 & $169$ & GIKMRVW & ELT & $0.53$ & $0.92$ & $2.4 \cdot 10^{2}$ & $5.9 \cdot 10^{2}$ & $1.2 \cdot 10^{3}$ & $4.1$ & $14$ & $41$\\
 & $171$ & KPR & AEGHNQST & $1.4$ & $0.62$ & $1 \cdot 10^{2}$ & $1.9 \cdot 10^{2}$ & $3.1 \cdot 10^{2}$ & $1.5$ & $4.5$ & $11$\\
\\
\multirow{3}{*}{PGT121}
 & $330$ & FHLRSY & Q & $0.12$ & $1.4$ & $7.4$ & $1.6 \cdot 10^{2}$ & $5.3 \cdot 10^{2}$ & $0.15$ & $3.6$ & $16$\\
 & $332$ & AENV & DIKT & $1.1$ & $1.2$ & $80$ & $1.8 \cdot 10^{2}$ & $3.2 \cdot 10^{2}$ & $1.3$ & $4.2$ & $11$\\
 & $334$ & DS & GN & $1$ & $2$ & $15$ & $80$ & $1.7 \cdot 10^{2}$ & $0.26$ & $1.8$ & $5.9$\\
\\
\multirow{5}{*}{PGT145}
 & $121$ & K & E & $1$ & $1$ & $34$ & $92$ & $1.7 \cdot 10^{2}$ & $0.58$ & $2.1$ & $5.8$\\
 & $160$ & N & KY & $0.4$ & $0.2$ & $1.4 \cdot 10^{2}$ & $2.5 \cdot 10^{2}$ & $4.4 \cdot 10^{2}$ & $2.1$ & $6.2$ & $15$\\
 & $162$ & ST & AIP & $1.2$ & $1.1$ & $6.8 \cdot 10^{2}$ & $1.2 \cdot 10^{3}$ & $2 \cdot 10^{3}$ & $11$ & $29$ & $73$\\
 & $166$ & KR & AEGST & $0.73$ & $0.58$ & $58$ & $1.2 \cdot 10^{2}$ & $2 \cdot 10^{2}$ & $0.9$ & $2.8$ & $7.2$\\
 & $169$ & GIKLTV & EMRW & $0.59$ & $1.4$ & $8.8$ & $80$ & $1.9 \cdot 10^{2}$ & $0.16$ & $1.8$ & $6.1$\\
\\
\multirow{5}{*}{PGT151}
 & $512$ & A & GT & $1.1$ & $0.57$ & $8.6 \cdot 10^{2}$ & $2.7 \cdot 10^{3}$ & $4.1 \cdot 10^{3}$ & $16$ & $59$ & $1.6 \cdot 10^{2}$\\
 & $611$ & GN & DS & $1.3$ & $2$ & $3.6 \cdot 10^{2}$ & $8.9 \cdot 10^{2}$ & $1.3 \cdot 10^{3}$ & $6.5$ & $20$ & $50$\\
 & $613$ & ST & CN & $0.28$ & $0.7$ & $1.2 \cdot 10^{3}$ & $2.8 \cdot 10^{3}$ & $3.5 \cdot 10^{3}$ & $21$ & $58$ & $1.4 \cdot 10^{2}$\\
 & $637$ & DN & EKST & $0.83$ & $0.41$ & $1.2 \cdot 10^{2}$ & $2.1 \cdot 10^{2}$ & $3.5 \cdot 10^{2}$ & $1.8$ & $5.1$ & $13$\\
 & $639$ & T & IM & $1$ & $1$ & $1.1 \cdot 10^{3}$ & $1.6 \cdot 10^{3}$ & $2.6 \cdot 10^{3}$ & $16$ & $42$ & $96$\\
\\
\multirow{5}{*}{VRC01}
 & $197$ & DN & S & $0.5$ & $1$ & $7.3 \cdot 10^{2}$ & $1.6 \cdot 10^{3}$ & $2.3 \cdot 10^{3}$ & $12$ & $35$ & $88$\\
 & $279$ & DN & AHK & $0.33$ & $0.22$ & $2 \cdot 10^{2}$ & $4.2 \cdot 10^{2}$ & $9.4 \cdot 10^{2}$ & $3.2$ & $10$ & $30$\\
 & $280$ & N & D & $1$ & $1$ & $4.4 \cdot 10^{2}$ & $1 \cdot 10^{3}$ & $1.8 \cdot 10^{3}$ & $7.3$ & $24$ & $63$\\
 & $281$ & AP & TV & $1.1$ & $1.1$ & $52$ & $1.3 \cdot 10^{2}$ & $2.5 \cdot 10^{2}$ & $0.9$ & $3.1$ & $8.6$\\
 & $458$ & G & D & $0.5$ & $1$ & $2.9 \cdot 10^{2}$ & $1.2 \cdot 10^{3}$ & $2.2 \cdot 10^{3}$ & $6$ & $26$ & $75$\\
\\
\multirow{5}{*}{VRC34}
 & $512$ & A & GT & $1.1$ & $0.57$ & $7.2 \cdot 10^{2}$ & $1.4 \cdot 10^{3}$ & $2.9 \cdot 10^{3}$ & $11$ & $34$ & $92$\\
 & $524$ & GP & AERS & $1.5$ & $0.67$ & $45$ & $95$ & $1.5 \cdot 10^{2}$ & $0.71$ & $2.2$ & $5.6$\\
 & $88$ & N & K & $0.26$ & $0.26$ & $5.7 \cdot 10^{2}$ & $1.4 \cdot 10^{3}$ & $1.9 \cdot 10^{3}$ & $9.8$ & $30$ & $74$\\
 & $90$ & ST & AEK & $0.48$ & $0.8$ & $1.4 \cdot 10^{3}$ & $2.5 \cdot 10^{3}$ & $3.3 \cdot 10^{3}$ & $19$ & $57$ & $1.3 \cdot 10^{2}$\\
\end{tabular}
} 
\caption{{\bf Selection and mutational target size for escape-mediating sites against each bNAb.} Shown are the sites (column 1) and the susceptible and the escape amino acids (column 2) for each bNAb. We called patterns for 10-1074 and CD4bs antibodies VRC01 and 3BNC117 using genetic trial data and the remainder using DMS data. The inferred mutational target size of escape at each site (forward $\mu$ and backward $\mu^\dagger$ mutation rates) is shown in column 3. 
The major quantiles ($10\%, \,50\%\, \text{(median)}, 90\%)$ associated with the inferred site-specific Bayesian posterior of the  scaled selection strength $\hat{\sigma} = \sigma/ \theta_\ts$  are shown in column 4. The corresponding quantiles for the strength of selection  $\sigma = \hat{\sigma} \theta_\ts$, after convolving the   posterior for the scaled selection $\hat{\sigma}$ with  the reservoir-corrected intra-patient diversity of HIV $\xi \theta$  are shown in column 5.  
}
\label{tab:full_antibody_data}
\end{table*}